\documentclass[aps,preprint,12pt]{revtex4}
\usepackage{hyperref}
\usepackage{graphicx}        
\def\N{{\cal N}}

\def\dslash{\ \lower 0.4ex\hbox{$\nabla$}\kern -1.0em\lower 0.2
ex\hbox{ $\slash$}\ }

\begin{document}

\title{Cosmological Entropy Bounds}
\author{Ram Brustein \\
Department of Physics, \\ Ben-Gurion University, \\  Beer-Sheva 84105, Israel \\
\texttt{ramyb@bgu.ac.il}}
%
%


\begin{center}
To appear in {\em  ``String theory and fundamental interactions"} \\
published in celebration of the 65$^{\rm th}$ birthday of Gabriele Veneziano \\
eds. M. Gasperini and J. Maharana \\
Lecture Notes in Physics, Springer Berlin/Heidelberg, 2007,
http://www.springerlink.com/content/1616-6361.
\end{center}

\begin{abstract}

I review some basic facts about entropy bounds in general and about cosmological
entropy bounds. Then I review the Causal Entropy Bound, the conditions for its
validity and its application to the study of cosmological singularities. This
article is based on joint work with Gabriele Veneziano and subsequent related
research.
\end{abstract}
\maketitle

\section{To Gabriele}
\label{sec:1}

On the occasion of your 65th birthday may you continue to find joy in science and
life as you have always had, and continue to help us understand our universe with
your creative passion and vast knowledge. It is a pleasure and an honor to
contribute to this volume and present one of the subjects among your many
interests. Thank you for explaining to me why entropy bounds are interesting and
for your collaboration on this and other subjects.

\section{Introduction}
\label{sec:2}

\subsection{What are entropy bounds?}

The second law of thermodynamics states that the entropy of a closed system tends
to grow towards its largest possible  value. But what is this  maximal value?
Entropy bounds aim to answer this question.

Bekenstein \cite{Bek1} has suggested that for a system of energy $E$ whose size $R$
is larger than its gravitational radius $R > R_g \equiv 2G_N E$, entropy is bounded
by
\begin{equation}
\label{beb1}
 S\le ER/\hbar = R_g~ R ~l_P^{-2}.
\end{equation}
Here $l_P$ is the Planck length. This is known as the Bekenstein entropy bound
(BEB).

Entropy bounds are closely related to black hole (BH) thermodynamics and their
interplay with their ``normal" environment. They are also probably associated with
instabilities to forming BH's, however, this has not been proved in an explicit
calculation. The original argument of Bekenstein was based on the Geroch process: a
thought experiment in which a small thermodynamic system is moved from infinity
into a BH. The small system is lowered slowly until it is just outside the BH
horizon, and then falls in. By requiring that the generalized second law (GSL) will
not be violated one gets inequality (\ref{beb1}).

A long debate about the relationship between entropy bounds and the GSL has been
going on. On one side  Unruh, Wald and others \cite{waldbeb1,waldbeb2} have argued
that the GSL holds automatically, so that entropy bounds cannot be inferred from
situations where the law seems to be violated. They argue that the microphysics
will eventually take care of any apparent violation. Consequently, they argued that
the BEB does not have to be postulated as a separate requirement in addition to the
GSL. Responding to their arguments Bekenstein  \cite{Bek2} has argued that it is
not always obvious in a particular example how the system avoids violating the
bound and analyzed in detail several of the purported counterexamples of this type
and demonstrated in each case the specific mechanism enforcing the bound.

Holography \cite{holography} (see below) suggests that the maximal entropy of any
system is bounded by $ S_{HOL}\le A l_P^{-2}$, where $A$ is the area of the
space-like surface enclosing a certain region of space. For systems of limited
gravity $R > R_g$, and since $A=R^2$, the BEB implies the holography bound. Physics
up to scales of about 1 TeV is very well described in terms of quantum field
theory, which uses, roughly, one quantum mechanical degree of freedom (DOF) for
each point in space (the number of DOF is the logarithm of the number of
independent quantum states). This seems to imply  that $ S(V) \sim V$, but the BEB
states that $ S(V) \le A$. The BEB does not seem to depend on the detailed
properties of the system and can thus be applied to any  volume $V$ of space in
which gravity is not dominant.  The bound is saturated by the Bekenstein-Hawking
entropy associated with a BH horizon, stating that no stable spherical system can
have a higher entropy than a BH of equal size.

A bold interpretation of the BEB was proposed by 't~Hooft and Susskind
\cite{holography}, that  the number of independent quantum DOF contained in a given
spatial volume $V$ is bounded by the surface area of the region. In a later
formulation by Bousso \cite{boussorev} their conjecture reads ``{\em a physical
system can be completely specified by data stored on its boundary without exceeding
a density of one bit per Planck area}". In this sense the world is two-dimensional
and not three-dimensional, for this reason their conjecture is called the
holographic principle. The holographic principle postulates an extreme reduction in
the complexity of physical systems, and is not manifest  in a description of nature
in terms of quantum field theories on curved space.  It is widely believed that
quantum gravity has to be formulated as a holographic theory.  This point of view
has received strong support from the ADS/CFT duality
\cite{Maldacena,Aharony:1999ti}, which defines quantum gravity non-perturbatively
in a certain class of space-times and involves only the physical degrees admitted
by holography.

One way of viewing entropy bounds is that they are new laws of nature that have to
supplement the equations that govern any fundamental theory of quantum gravity.
From this perspective the entropy bounds and the holographic principle are presumed
to be valid for any physical system and their ``true" form has to be unravelled. An
alternative perspective is that entropy bounds will be automatically obeyed by any
physical system and will be a consequence of the fundamental dynamical equations.
As such entropy bounds will not provide additional independent constraints on the
system's evolution. In the final fundamental theory entropy bounds will be
tautologically correct. My personal view on this issue at the present time is
closer to the second point of view.

My current perspective is that without detailed knowledge of the dynamical
equations that govern physics at the shortest distance scales and at the highest
energies it is hard to make detailed quantitative use of entropy bounds. They are
very useful as qualitative tools in the absence of the final fundamental theory of
quantum gravity when one is trying to determine whether a candidate theory is
correct by studying its consequences. As I will explain they are particulary useful
in discriminating among cosmologies that are suspect of being unphysical for
various reasons.

\subsection{What are cosmological entropy bounds?}

Is it possible to extend entropy bounds to more general situations, for example, to
cosmology? In 1989 Bekenstein proposed \cite{Bek3} that it might be possible to
apply the BEB to a region as large as the particle horizon $d_p$: $ d_p(t) = a(t)
\int_{t_{\rm initial}}^t dt'/a(t')$, $a(t)$ being the scale factor of an
Friedman-Robertson-Walker (FRW) universe. If the entropy of a visible part of the
universe obeys the usual entropy bound from nearly flat space situations, then
Bekenstein suggested that the temperature of the universe is bounded and therefore
certain cosmological singularities are avoided. The proposal to apply the
holographic bound from nearly flat space to cosmology was first made by Fischler
and Susskind \cite{FS} and later extended and modified by Bousso \cite{boussorev}.
E. Verlinde \cite{EV} proposed an entirely holographic bound on entropy stating
that the subextensive component of the entropy (the ``Casimir entropy") of a closed
universe has to be less than the entropy of a BH of the same size.

To appreciate the necessity to modify the BEB in some situations let us think
\cite{Branepuzzle} about a  box of relativistic gas in thermal equilibrium at a
temperature $T$. We assume that the gas consists of $\N$ independent DOF and is
confined to a box of macroscopic linear size $R$. We further assume that $R$ is
larger than any fundamental length scale in the system, and in particular that $R$
is much larger than the Planck length $R\gg l_P$. The volume of the box is $V=R^3$.
Since the gas is in thermal equilibrium its energy density is $\rho=\N T^4$ and its
entropy density is $s=\N T^3$ (here and in the following we systematically neglect
numerical factors). Here we are interested in the case $RT>1$ which means that the
size of the box is larger than the thermal wavelength $1/T$. The case $RT<1$ has
been considered previously in \cite{shortest}. In this case the temperature is not
relevant, rather the field theory cutoff $\Lambda$ was shown to be the relevant
scale.

Under what conditions is this relativistic gas unstable to the creation of BH's?
The simplest criterion which may be used to determine whether an instability is
present is a comparison of the total energy in the box $E_{\rm Th}=\N T^4 R^3$ to
the energy of a BH of the same size $E_{\rm BH}=M_P^2 R$ ($M_P$ is the Planck
mass). The two energies are equal when $T^4= 1/\N M_P^2/R^2$. So thermal radiation
in a box has a lower energy than a BH of the same size if
\begin{equation}
\label{energy} (TR)^4 < \frac{1}{\N} M_P^2  R^2.
\end{equation}
Another way to determine the presence of an instability to creation of BH's  is to
compare the thermal entropy $S_{\rm Th}=\N T^3 R^3$ to the entropy of the BH
$S_{\rm BH}=M_P^2 R^2$. They are equal when $T^3= 1/\N M_P^2/R$. So thermal
radiation in a box has a lower entropy than  a BH of the same size if
\begin{equation}
\label{entropy}
 (TR)^3 < \frac{1}{\N} M_P^2 R^2.
\end{equation}
From eqs. (\ref{energy}) and (\ref{entropy}) it is possible to conclude the well
known fact that for fixed $R$ and $\N$, if the temperature is low enough the
average thermal free energy is not sufficient to form BH's. For low temperatures
the thermal fluctuations are weak and they do not alter the conclusion
qualitatively.

Now imagine raising the temperature of the radiation from some low value for which
conditions (\ref{energy}), (\ref{entropy}) are comfortably satisfied to higher and
higher values such that eventually condition (\ref{energy}) is saturated. Since
$TR>1$ eq.~(\ref{energy}) is saturated before eq.~(\ref{entropy}). We assume that
the size of the box $R$ is fixed during this process (recall that the number of
species $\N$ is also fixed), and estimate the backreaction of the radiation energy
density on the geometry of the box to determine whether the assumption that the
geometry of box is fixed is consistent. To obtain a simple estimate we assume that
the box is spherical, homogeneous and isotropic. Then its expansion or contraction
rate is given by the Hubble parameter $H= \dot R/R$, which is determined by the
$00$ Einstein equation $H^2 M_{P}^2=\N T^{4}$. However, if eq.~(\ref{energy}) is
satisfied then $\frac{1}{R^2} M_{P}^2=\N T^{4}$, and therefore $HR\sim 1$. The
conclusion is that if eq.~(\ref{energy}) is saturated then the gravitational time
scale is comparable to the light crossing time of the box, and therefore it is
inconsistent to assume that the box has a fixed size which is independent of the
energy density inside it.

Thus we have shown that  it is  not possible to ignore the backreaction of the gas
on the geometry under all circumstances. Sometimes the backreaction has to be taken
into account. When the BEB is near saturation we have found that the basic
assumptions have to be changed so it has to be modified to adapt to an
intrinsically time dependent situation.

\subsection{Why is it reasonable to expect cosmological entropy bounds?}

Some have argued incorrectly that it is impossible to discuss entropy bounds in
cosmology. They argue that the universe is the whole system and thus one cannot
apply thermodynamical arguments that sometimes rely on separating a subsystem from
a heat reservoir. This argument is false as the following braneworld thought
experiment explicitly demonstrated \cite{Branepuzzle}. Let us consider a brane
moving in a higher dimensional BH background. From the brane point of view it
experiences a cosmological evolution and one can imagine that the brane falls into
the BH and disappears from an external observer's  view into the BH horizon. We are
thus in a situation similar to the one envisaged in the Geroch process: the thought
experiment in which a thermodynamic system is absorbed by a BH. The aim is to
design the process such that the energy absorbed by the BH is minimal. In such a
way the entropy that the BH gains will also be minimal, as both the energy and the
entropy of the BH depend only on its mass after the absorption. We can make the
entropy balance during the process and see under which conditions the GSL is
respected.

We can gain some insight by modelling a 4D radiation-dominated (RD) universe as a
brane moving in an AdS$_{5}$-Schwarzschild spacetime. For the BH in AdS to be the
dominant configuration over an AdS space filled with thermal radiation as required
for our analysis to be relevant, The BH must be large and hot compared to the
surrounding AdS$_{5}$ \cite{Witten}. In this limit the closed 4D universe can be
treated as flat. The motion of the brane through the bulk spacetime is viewed by a
brane observer as a cosmological evolution. According to the prescription of the RS
II model \cite{Randall:1999vf}, the 4D brane is placed at the $Z_2$ symmetric point
of the orbifold. On the other hand, in the so called mirage cosmology
\cite{Kraus:1999it,Kehagias:1999vr}, the brane is treated as a test object
following a geodesic motion. In both cases the evolution of the brane in the
AdS$_{5}$-Schwarzschild bulk mimics an FRW RD cosmology. From the 5D perspective
one may expect some limits on the entropy of the brane by considering what happens
when the BH swallows the brane.

\subsection{What are cosmological entropy bounds good for?}

Our interest in entropy bounds in general and cosmological entropy bounds in
particular originated from the interest in determining the fate of cosmological
singularities. Specifically, we were interested in finding whether the bounce that
is an essential part of the pre-big-bang (PBB) scenario of string cosmology
\cite{PBBrev} can be physically realized or perhaps there is some principle that
requires the solution to be singular. We needed a general principle because string
theory could not provide an explicit enough model of the hypothetical bounce
transition. The traditional tools for finding such criteria were the energy
conditions that are used in the singularity theorems. However, the use of energy
conditions is limited because there are examples of cosmologies that do not seem to
be problematic in any of their physical properties and for which the singularity
theorems are not applicable because some of the energy conditions are violated. On
the other hand, there are examples of cosmologies for which we expect some problems
while the singularity theorems seem perfectly valid.

Let us consider, for example, the scale factor for a closed deSitter Universe. This
is  a closed Universe containing a positive cosmological constant $\Lambda$. In
$D=4$ is given by
$a(t)=(\frac{\Lambda}{3})^{-1/2}\cosh{\sqrt{\frac{\Lambda}{3}}t}$, showing a bounce
at $t=0$. The bounce is not allowed by the classic singularity theorems. This is
not surprising since the sources of this model violate the strong energy conditions
(SEC). The reliability of the SEC as a criterion of discriminating physical and
unphysical solutions is therefore questionable (as is well known in the context of
inflationary cosmology). Conversely, in a 4D contracting universe filled with
radiation consisting of $\N$ species in thermal equilibrium, the singularity
theorems imply the the solution will reach a future singularity. But entropy bounds
indicate expected problems already when $T\sim M_{\rm P}/\N^{1/2}$ as we will show
later.

\section{The Causal Entropy Bound}
\label{sec:3}

\subsection{The Hubble Entropy Bound}
\label{sec:21}

Motivated by the necessity to resolve the apparent singularity in the lowest order
classical PBB scenario Veneziano has studied the possible role of entropy bounds
and proposed the Hubble entropy bound (HEB) \cite{HEB}. The physical motivations
leading to the proposal of the HEB are $(i)$ that in a given region of space the
entropy is maximized  by the largest BH that can fit in it; $(ii)$ that the largest
BH that can hold together without falling apart in a cosmological background has
typically the size of the Hubble radius. In the following we review the basic ideas
that led to Veneziano's proposal of the HEB.

Veneziano considered the possibility that the BEB or holography bounds can be
applied to an arbitrary sphere of radius $R$, cut out of a homogeneous cosmological
space. Entropy in cosmology is extensive so it grows like $R^3$, but the boundary's
area grows like $R^2$. Hence, at sufficiently large $R$, the (naive) holography
bound must be violated. On the other hand, $S_{BEB} \sim E R \sim R^4$ appears to
be safer at large $R$.

In order to show how inadequate the naive bounds are in cosmology, Veneziano
applied them at the Planck time $t \sim t_P \sim 10^{-43}~{\rm s}$, within standard
FRW cosmology, to the region of space that has become our visible Universe today.
The size of that region at $t \sim t_P$ was about $10^{30}$ in units of the Planck
length $l_P$, and the entropy density was of about Planckian. Thus, the actual
entropy of the patch is
\begin{eqnarray}
&&S \sim (10^{30})^3 = 10^{90}
\end{eqnarray}
while
\begin{eqnarray}
  &&S_{BEB} \sim \rho R^4/\hbar \sim R^4/l_P^4 \sim 10^{120} \;\;,
 S_{HOL} \sim R^2 l_P^{-2} \sim 10^{60}\; .
\end{eqnarray}
The actual entropy lies at the geometric mean between the two naive bounds, making
one false and the other quite useless. The two bounds differ by a factor $(H
d_p)^2$. While such a factor is of order unity in FRW-type cosmologies, it can be
huge after a long period of inflation. For this reason the  (naive) holographic
entropy bound appears to be stronger than the cosmological version of the BEB, just
the opposite of what we argued to be the case for systems of limited gravity.

A sufficiently homogeneous Universe has a local time-dependent Hubble expansion
rate defined, in the synchronous gauge, by $
  H = \frac{1}{6}\partial_t~(\log \det g_{ij} )
$.  If $H$ does not vary much over distances $\sim 1/H$ then the Hubble radius
$1/H$ corresponds to the scale of causal connection. If on top of this homogeneous
background some isolated lumps of size much smaller that $1/H$ exist, then the
expansion of the Universe is irrelevant and the situation should be similar to that
of nearly flat space. Veneziano argued that it is possible in this case that a
single Hubble patch contains several BH's. The BH can coalesce and in the process
their entropy will increase. He argued further that this way of increasing entropy
has some limit since it is hard to imagine that a BH of size larger than $1/H$ can
form. The different parts of its horizon would be unable to  hold together. Strong
arguments in this direction were given long ago in the literature \cite{ch}. Thus,
the largest entropy in a region of space larger than $1/H$ is the one corresponding
to one BH per Hubble volume $1/H^{3}$. Using the Bekenstein--Hawking formula for
the entropy of a BH of size $1/H$ leads to the proposal of a ``Hubble entropy
bound", that the entropy is bounded by $S_{HEB}\equiv n_H S^H$, where $n_H$ is the
number of Hubble-size regions within the volume $V$, each one carrying maximal
entropy $S^H = l_P^{-2}  H^{-2}$,
\begin{equation}
S(V) < S_{HEB} \equiv  n_H S^H = V H^3 l_P^{-2}  H^{-2} = V H l_P^{-2} ~~.
\label{HB}
\end{equation}

The HEB is partly holographic since $S^H$ scales as an area, and partly extensive
since  $n_H$ scales as the volume. If  the HEB is applied to a region of size $d_p$
then the bound is the geometric mean of the BEB and the naive holography bound,
\begin{equation}
S_{HEB} = d_p^3 H l_P^{-2} = S_{BEB}^{1/2}~S_{HOL}^{1/2}~~. \label{HBvsBFS}
\end{equation}

\subsection{The Causal Entropy Bound}
\label{sec:22}

The Causal Entropy Bound (CEB) \cite{CEB} aims to improve the HEB. It is a
covariant bound applicable to entropy on space-like hypersurfaces. We do not
insist, a priori, on a holographic bound, but aim at generality of the hypersurface
and then investigate how holography may or may not work. For systems of limited
gravity Bekenstein's bound is the tightest bound, while, in other situations, the
CEB is the strongest one which does not lead to contradictions for space-like
regions.

We shall refer to entropy in a region as to a quantity proportional to the number
of DOF in that region. To be more precise, we shall exclude from consideration
entropy associated with the background gravitational field itself. We will however
take into account the entropy of the perturbations of the gravitational field. Let
us first state our proposal, and then motivate and test it. Consider a generic
spacelike hypersurface, defined by the equation $\tau = 0$, and a compact region
lying within it defined by $\sigma \le 0$. We have proposed that the entropy
contained in this region, $S(\tau = 0, \; \sigma \le 0)$, is bounded by $S_{CEB}$,
\begin{eqnarray}
&&S_{CEB}= l_P^{-2} \int\limits_{\sigma <0} d^4 x \sqrt{-g} \delta(\tau) \sqrt{
{\rm Max}_\pm\left[ (G_{\mu\nu}\pm R_{\mu\nu})
\partial^{\mu} \tau \partial^{\nu} \tau \right]}~ =
\nonumber \\ && l_P^{-1} \hbar^{-1/2} \int\limits_{\sigma <0} d^4 x \sqrt{-g}
\delta(\tau) \sqrt{ {\rm Max}_\pm \left[( T_{\mu\nu} \pm T_{\mu\nu} \mp {1\over2}
g_{\mu\nu}~T)
\partial^{\mu} \tau \partial^{\nu} \tau \right]}.
\label{CCB}
\end{eqnarray}
Here $G_{\mu\nu}$, $R_{\mu\nu}$ are the Einstein and Ricci tensor, respectively,
$T_{\mu\nu}$ is the energy-momentum tensor, and $T$ its trace. To derive the second
equality we have used Einstein's equations, $G_{\mu\nu} = 8 \pi G_N T_{\mu\nu}$.
Note the appearance of the square-root of the energy contained in the region and
that (\ref{CCB}) is manifestly covariant, and invariant under reparametrization of
the hypersurface equation: such an invariance requires a square-root of
$\partial^{\mu} \tau
\partial^{\nu} \tau $. Reality of $S_{CEB}$ is assured if sources
obey the weak energy condition, $T_{\mu\nu}\partial^{\mu} \tau
\partial^{\nu} \tau \ge 0$, since then the sum of the two
combinations in (\ref{CCB}), and thus their maximum, are positive. The weak energy
condition is sufficient but not necessary for reality. We expect that for physical
systems reality will be always guaranteed.

Since eq.~(\ref{CCB}) applies to any space-like region, it can be written in a
local form rather than in an integrated form by introducing an entropy current
$s_{\mu}$ such that $S = \int d^4 x \sqrt{-g}\delta(\tau) s_{\mu}
\partial^{\mu} \tau $.
Then (\ref{CCB}) becomes equivalent to ($\lambda^{\mu}$ being an arbitrary
time-like vector):
\begin{equation}
s_\mu \lambda^\mu \le l_P^{-1} \hbar^{-1/2} \sqrt{ {\rm Max}_\pm \left[( T_{\mu\nu}
\pm T_{\mu\nu} \mp {1\over2} g_{\mu\nu}~T) \lambda^{\mu} \lambda^{\nu}\right]}.
\label{CCBdiff}
\end{equation}

In the limit in which the hypersurface is lightlike, $\partial^{\mu} \tau
\partial_{\mu} \tau = 0$, eqs.~(\ref{CCB}), (\ref{CCBdiff}) read:
\begin{eqnarray}
S_{CEB} &=& \int_{\sigma <0} d^4 x \sqrt{-g} \delta(\tau) \sqrt{T_{\mu\nu}
\partial^{\mu} \tau \partial^{\nu} \tau} \;,
\nonumber \\ s_\mu \lambda^\mu &\le& l_P^{-1} \hbar^{-1/2} \sqrt{ T_{\mu\nu}
\lambda^{\mu} \lambda^{\nu} } \;, \;\; \lambda_{\mu} \lambda^{\mu} = 0 \; ,
\label{CCBnull}
\end{eqnarray}
and become closely related to the assumptions made in \cite{Wald} (eq.~(1.10)). We
already see signs here that the physics at short scales and high energies is
important in determining the value of the maximal entropy because $T_{\mu\nu}$ is
generically at least quadratic in the fields.

The physical motivations leading us to the above proposal are similar to those used
to motivate the HEB: $(i)$ that entropy is maximized, in a given region of space,
by the largest BH that can fit in it; $(ii)$ that the largest BH that can hold
together without falling apart in a cosmological background has typically the size
of the Hubble radius. The second assumption clearly needs to be refined and,
possibly, to be defined covariantly. With such a goal in mind, we will proceed as
follows: we will start by identifying a critical (``Jeans") length scale above
which perturbations are causally disconnected so that BH of larger size, very
likely, cannot form. We will first find this causal connection (CC) scale $R_{CC}$
for the simplest cosmological backgrounds, then extend it to more general cases
and, finally, guess the completely general expression using general covariance.

In order to identify the CC scale for a homogeneous, isotropic and spatially flat
background, let us consider a generic perturbation around such a background in the
hamiltonian approach developed in \cite{BMV}. The Fourier components of the
(normalized) perturbation and of its (normalized) conjugate momentum satisfy
Schroedinger-like equations ${\widehat\Psi}_k{''}\!\!+\!\!\left[k^2-(S^{1/2}){''}
S^{-1/2}\right]{\widehat\Psi}_k\!\!\!=\!\!\!0$, ${\widehat\Pi}_k{''}\!\!+\!\!
\left[k^2-(S^{-1/2}){''} S^{1/2}\right] {\widehat\Pi}_k\!\!\!=\!\!\!0$, where $k$
is the comoving momentum, a prime denotes differentiation w.r.t. conformal time
$\eta$, and $S^{1/2}$ is the so-called ``pump field", a combination of the various
backgrounds which depends on the specific perturbation under study. The
perturbation equations clearly identify a ``Jeans-like" CC comoving momentum
\begin{eqnarray}
k_{CC}^2 &=& {\rm Max} \left[ (S^{1/2}){''} S^{-1/2} ~, ~(S^{-1/2}){''} S^{1/2} ~
\right] \nonumber \\ &=& {\rm Max} \left[ {\cal K}' + {\cal K}^2 ~, ~ - {\cal K}' +
{\cal K}^2 \right], \label{kjeans}
\end{eqnarray}
where ${\cal K}=(S^{1/2}){'} S^{-1/2}$. Equation~(\ref{kjeans}) always defines a
real $k_{CC}$ since the sum of the two quantities appearing on the r.h.s. is
positive semidefinite. Since tensor perturbations are always present, let us
restrict our attention to them. The ``pump field" $S^{1/2}$ is simply given, in
this case, by the scale factor $a(\eta)$ so that ${\cal K} \rightarrow {\cal H} =
a'/a$. Equation (\ref{kjeans}) is immediately converted into the definition of a
proper ``Jeans" CC length $R_{CC} = a k_{CC}^{-1}$. Substituting into
eq.~(\ref{kjeans}), and expressing the result in terms of proper-time quantities,
we obtain (for tensor perturbations) $ R_{CC}^{-2} = {\rm Max} \left[ \dot{H} + 2
H^2~, ~ - \dot{H}~\right]. $ Before trying to recast this equation in a more
covariant form let us remove the assumption of spatial flatness by introducing the
usual spatial-curvature parameter $\kappa$ ($\kappa = 0, \pm1$). The study of
perturbations in non-flat space is considerably more complicated than in a
spatially-flat background. The final result, however, appears to be extremely
simple \cite{Garriga,GPV}, and can be obtained from the flat case by the following
replacements in eq.~(\ref{kjeans}): ${\cal H}^2 \rightarrow {\cal H}^2 + \kappa$,
${\cal H}' \rightarrow {\cal H}' $. Using this simple rule  we arrive at the
following generalization
\begin{equation}
R_{CC}^{-2} = {\rm Max} \left[ \dot{H} + 2 H^2~ + \kappa/a^2, ~ - \dot{H}~ +
\kappa/a^2 \right]. \label{RCCk}
\end{equation}

At this point we could have introduced anisotropy in our homogeneous background and
study perturbations with or without spatial curvature. Instead, we adopt a shortcut
route. We observe that the $00$ components of the Ricci and Einstein tensors for
our background are given by $ R_{00} = -3 (\dot{H} + H^2) ~, ~ G_{00} = 3 ( H^2 +
\kappa/a^2)\;. $ Obviously,
\begin{eqnarray}
R_{CC}^{-2} &=& {1\over 3}{\rm Max}_{\mp} \left(G_{00} \mp R_{00} \right) \nonumber
\\ &=& 4 \pi G_N ~{\rm Max} \left[{\rho \over 3} - p~,~ \rho + p \right] \;,
\label{RCCC}
\end{eqnarray}
where we have inserted Einstein's equations using as an example a perfect-fluid
energy momentum tensor $T^\mu_{\ \nu}=diag(\rho,-p,-p,-p)$. Equation~(\ref{RCCC})
is guaranteed to define a real $R_{CC}$ if the weak energy condition (reading here
$\rho>0$) holds, since the sum of the two combinations is positive in this case. In
general, other perturbations may compete with tensor perturbations and define a
smaller $R_{CC}$. In this case, the symbol $~{\rm Max}$ in the above equations also
applies to the various types of perturbations. This may help to ensure reality of
$R_{CC}$ in all physical situations.

As a final step, let us convert eq.~(\ref{RCCC}) into an explicitly covariant bound
on entropy.  Using $R_{CC}$ as the maximal scale for BH's, we get a bound on
entropy which scales like $ S \sim V R_{CC}^{-3} ~ R_{CC}^2 l_P^{-2}~ = ~ V
R_{CC}^{-1} ~ l_P^{-2}. $ We now express $R_{CC}^{-1}$ as in (\ref{RCCC}) in terms
of the components of the Ricci and Einstein tensors in the direction orthogonal to
the hypersurface on which the entropy is being computed. This can be done
covariantly by defining the hypersurface through the equation $\tau =0$ and by
identifying the normal with the vector $\nabla^{\mu} \tau$. This procedure leads
immediately to the proposal (\ref{CCB}). The local form (\ref{CCBdiff}) clearly
follows by shrinking the space-like region to a point. Alternatively, using
standard $3+1$ ADM formalism \cite{ADM}, we can express the relevant components of
the Ricci and Einstein tensors in terms of the intrinsic and extrinsic curvature of
the hypersurface under study and arrive at the following final formula:
\begin{eqnarray}
 S_{CEB} &=& l_P^{-2}
 ~\int d^3 x \sqrt{h} \;
\left[{\rm Max}\left(P~,~Q \right)\right]^{1/2}~, \label{CCBADM}
\end{eqnarray}
 where $P = {1 \over 2} {\cal R} +
\dot{\theta}
 +{2 \over 3} \theta^2 + \sigma^2 -
{\cal A}~, Q = {1 \over 2} {\cal R} - \dot{\theta} - 3 \sigma^2 + {\cal A}$. Using
standard notations, we have denoted by ${\cal R}$ the intrinsic 3-curvature scalar,
by $\theta$ the expansion rate, by $\sigma$ the shear, and by ${\cal A}$ the
``acceleration" given (for vanishing shifts $N_i$)in terms of the lapse function
$N$ by ${\cal A} = N^{-1} N^{,i}_{\ ;i}$.

\subsection{The CEB in D dimensions}

In order to generalize the CEB to arbitrary dimension $D$ \cite{BFV} we generalize
the causal-connection scale $R_{CC}$  by looking at perturbation equations in $D$
dimensions. For gravitons, in the case of flat universe, one finds \cite{GG}
\begin{eqnarray}\label{rccfl}
R_{CC}^{-2}=\frac{D-2}{2}{\rm Max}\left[\dot{H} + \frac{D}{2}H^2\, , -\dot{H} +
\frac{D-4}{2}H^2\right]\, .
\end{eqnarray}
If $H\gg\dot{H}$, $R_{CC} \propto H^{-1}$ and one recovers  HEB with a
$D$-dependent prefactor scaling as$\sqrt{D (D-2)}$. The above result generalizes to
the case of a spatially curved universe as we have explained previously,
\begin{eqnarray}
\label{rccgen} R_{CC}^{-2}=\frac{D-2}{2}{\rm Max}\left[\dot{H} + \frac{D}{2}H^2 +
\frac{D-2}{2}\frac{\kappa}{a^2}\, , -\dot{H} + \frac{D-4}{2}H^2 +
\frac{D-2}{2}\frac{\kappa}{a^2}\right]\!.\ \ \ \ \
\end{eqnarray}
A covariant definition of  $R_{CC}$ is obtained by expressing (\ref{rccgen}) in
terms of the $00$ components of curvature tensors. We find
\begin{eqnarray}
 \label{rccgen1}
R_{CC}^{-2}=\frac{D-2}{2(D-1)}{\rm Max}\left[G_{00}\mp R_{00}\right] =4\pi G_{N}
\left[\frac{1}{D-1}\rho-p\, , \frac{2D-5}{D-1}\rho+p\right]\!,\ \ \ \ \
\end{eqnarray}
where, to derive the second equality, we have used Einstein's equations,
$G_{\mu\nu} = 8 \pi G_N T_{\mu\nu}$ and a perfect-fluid form for the
energy-momentum tensor.

The Bekenstein-Hawking entropy  of a Schwarzchild BH of radius $R_{BH}$ in $D$
dimensions is given by $S={\cal A}/4 l_P^{D-2}$. The generalization of $S_{\rm
CEB}$ for a region of proper volume $V$ is therefore
\begin{eqnarray}
\label{scebD} S_{\rm CEB}=\beta n_H S^{BH}=\beta \frac{V}{V(R_{CC})} \frac{{\cal
A}}{4 l_P^{D-2}}
\end{eqnarray}
where $n_H \equiv \frac{V}{V(R_{CC})}$ is the number of causally connected regions
in the volume considered,  $V(x)$ denotes the volume of a region of size $x$, and
$\beta$ is a fudge factor reflecting current uncertainty on the actual limiting
size for BH stability. For a spherical volume in flat space we have
$V(x)=\Omega_{D-2} x^{D-1}/(D-1)$, with $\Omega_{D-2}=2\pi^{(D-1)/2}/
\Gamma\left(\frac{D-1}{2}\right)$. But in general the result is different and
depends on the spatial-curvature radius.

Following Ref.~\cite{CEB}, the expression for $S_{\rm CEB}$ in $D$ dimensions can
be rewritten in the explicitly covariant form
\begin{eqnarray}
\label{scebB} &&S_{\rm CEB}= B  l_P^{-(D-2)}\int\limits_{\sigma<0}d^Dx\
\sqrt{-g}\delta(\tau) \sqrt{{\rm Max}_{\pm}[(G_{\mu\nu}\pm R_{\mu\nu})
\partial^{\mu}\tau\partial^{\nu}\tau]} =
\nonumber \\ &&\hspace{-.4in} B (8 \pi)^{1/2} l_P^{-D/2 +1} \int\limits_{\sigma <0}
d^Dx\ \sqrt{-g} \delta(\tau) \sqrt{ {\rm Max}_\pm \left[( T_{\mu\nu} \pm T_{\mu\nu}
\mp {1\over2} g_{\mu\nu}~T)
\partial^{\mu} \tau \partial^{\nu} \tau \right]}\!,\ \ \ \
\end{eqnarray}
where $\sigma <0$ defines the spatial region inside the $\tau = 0$ hypersurface
whose entropy we are discussing, and $T$ is the trace of the energy-momentum
tensor.

The prefactor $B$ can be fixed by comparing eqs.~(\ref{scebD}) and (\ref{scebB}).
Let us consider the expression (\ref{scebD}) in the limit $R_{CC}\ll a$, where $a$
is the radius of the Universe. In this case, over a region of size $R_{CC}$ we may
neglect spatial curvature and write $V(R_{CC}) = \Omega_{D-2} R_{CC}^{D-1}/(D-1)$,
and the area of the BH horizon as ${\cal A}=\Omega_{D-2} R_{BH}^{D-2}$, thus giving
(apart for negligible terms of order $(R_{CC}/a)^2$)
\begin{eqnarray}
 \label{cebd1}
S_{\rm CEB} = \beta \frac{D-1}{4} V R_{CC}^{-1} l_P^{-(D-2)} = B
~\sqrt{\frac{2(D-1)}{D-2}}~ V R_{CC}^{-1} l_P^{-(D-2)}  \, .
\end{eqnarray}
This fixes $B =  \sqrt{\frac{(D-1)(D-2)}{32}} \beta$.

Since (\ref{scebB}) applies to any space-like region, it can be rewritten in a
local form as in a 4D case by introducing an entropy current $s_{\mu}$ such that $S
= \int d^D x \sqrt{-g}\delta(\tau) s_{\mu}
\partial^{\mu} \tau $.
Then (\ref{scebB}) becomes equivalent to (with $\lambda^{\mu}$ a arbitrary
time-like vector):
\begin{equation}
\label{CCBdiffD} s_\mu \lambda^\mu \le l_P^{-D/2 +1} (8 \pi)^{1/2} B \sqrt{ {\rm
Max}_\pm \left[( T_{\mu\nu} \pm T_{\mu\nu} \mp {1\over2} g_{\mu\nu}~T)
\lambda^{\mu} \lambda^{\nu}\right]}\, .
\end{equation}

In the limit of a light-like vector $\lambda$ we get one of the conditions proposed
by Flanagan et al. \cite{Wald}  in order to recover Bousso's proposal. Their bound
corresponds (in $D=4$) to $B = \frac{1}{4\pi}$ and  could be used to fix $\beta$
(assuming that it is $D$-independent).

For systems of limited gravity the BEB is tighter than the CEB, $S_{BEB}<S_{CEB}$.
Therefore, in all systems for which the BEB is obeyed, the CEB will be obeyed as
well. hence our bound is most interesting for systems of strong gravity, and in
particular in cosmology.

For general collapsing regions we have limited computational power. While the local
form (\ref{CCBdiff}) looks most appropriate for the study of collapsing regions,
most likely the analysis of the general case will need the use of numerical
methods. We can qualitatively check cases that are similar to the cosmological ones
\cite{MTW2}, such as homogeneous, isotropic contracting pressureless regions, or a
contracting homogeneous, isotropic region filled with a perfect fluid. The
pressureless case can be described by a Friedman interior and a Schwarzschild
exterior. Since CEB is valid for the analogue cosmological solution it is also
valid for this case.

A particularly interesting case is that of the (generically non-homogeneous)
collapse of a stiff fluid ($p = \rho$) which can be mapped by a simple field
redefinition onto the dilaton-driven inflation of string cosmology \cite{PBBrev}.
In this case one finds a constant $S_{CEB}$ in agreement with the HEB result
\cite{HEB}. Hence, no problem arises in this case, even if one starts from a
saturated $S_{CEB}$ at the onset of collapse. For non-stiff equations of state, the
situation appears less safe if one starts near saturation. However, care must be
taken in this case of perturbations which tend to grow non linear and form
singularities on rather short time scales. Such cases cannot be described
analytically, but have been looked at numerically.

\subsection{The CEB in cosmology}

The universe is a system of strong self-gravity. The geometry of the universe is
determined by self-gravity, and the size of the universe is at least its
gravitational radius. The strongest challenges to entropy bounds in general, and to
the CEB in particular, come from considering (re)collapsing universes.

In homogeneous and isotropic $D$ dimensional cosmological backgrounds we have found
the dependence of $R_{\rm CC}$ on the Hubble parameter $H(t)$, its time-derivative
$\dot{H}(t)$, and the scale factor $a(t)$ in eqs.~(\ref{rccgen}), (\ref{rccgen1}),
\begin{eqnarray}
\label{rcch} R_{\rm CC}^{-2}&=&\frac{D-2}{2}{\rm Max}\left[ \dot{H} +
\frac{D}{2}H^2 + \frac{D-2}{2}\frac{\kappa}{a^2}, -\dot{H} +
\frac{D-4}{2}H^2 + \frac{D-2}{2}\frac{\kappa}{a^2}\right] \nonumber \\
&=&\frac{4\pi G_{\rm N}}{D-1} {\rm Max} \biggl[
 \rho - (D-1)p\, , (2D-5)\rho + (D-1)p
 \biggr],
\end{eqnarray}
where $\kappa=0,\pm1$ determines the spatial curvature.  Notice that $R_{\rm CC}$
is well defined if $\rho$ is positive because the maximum in eq.~(\ref{rcch}) is
larger than the average of the two entries in the brackets, and the average is
equal to $2(D-2)\rho$.

The following four cases exhaust all possible types of cosmologies \cite{CEB,BFM}:
\begin{enumerate}

\item $|\dot H|\sim H^2 \sim |k|/a^2$, or $|\dot H|\sim H^2 \gg |k|/a^2$.
  In this case effective energy density and
  pressure are of the same order, $\rho\sim p$. All length scales
  that may be considered in entropy bounds, such as particle horizon,
  apparent horizon, $R_{\rm CC}$ and the Hubble radius are
  parametrically equal. This case includes non-inflationary FRW
  universes with matter and radiation.

\item $H^2 \gg |k|/a^2,|\dot H|$. In this case $|\rho+ p| \ll \rho$, and the
  universe is inflationary. In this case $R_{CC}$ is parametrically equal to $|H|^{-1}$.

 \item $|\dot H|\gg H^2,\ |k|/a^2$. In this
case \ $|\rho| \ll p$.  Since $\rho$
  and $p$ are the effective energy density and pressure, there are no
  problems with causality. This case occurs, for instance, near the
  turning point of an expanding universe which recollapses, or
  near a bounce of a contracting universe which reexpands.

\item $k/a^2 \gg |\dot H|,\ H^2$. In this case the spatial curvature
  determines the causal connection scale. This occurs, for example,
  when both $H$ and $\dot H$ vanish as in a closed Einstein Universe.

\end{enumerate}

We will first describe several cosmological models and explain how they satisfy the
CEB. Then we will present in a general form the conditions on sources that
guarantee the validity of the CEB.

\subsubsection{A radiation dominated Universe}

\begin{figure}
\centering
\includegraphics[height=4cm]{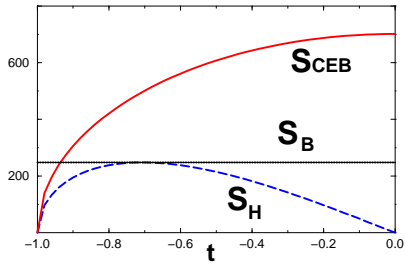}
\caption{ {\label{fig1}} $S_{CEB}$ compared with $S_{H}= (D-2) \frac{HV}{4G_N}$ and
$S_B \equiv  2 \pi R E /(D-1)$ in the expanding phase of a closed $D=4$, RD
Universe. Here we set $\beta=\frac{D-2}{D-1}$.}
\end{figure}

Our first example is a radiation dominated universe in $D$ dimensions. In this case
$\rho = (D-1) p$ and the $00$ equation for the scale factor is
\begin{eqnarray}
H^2 +\frac{ \kappa}{ a^2}=\frac{16\pi G_{N}}{(D-1)(D-2)}\rho = \frac{16\pi
G_{N}}{(D-1)(D-2)}\rho_{0}R_{0}^D a^{-D},  \kappa=\pm1,0. \ \ \
\end{eqnarray}
In terms of the conveniently rescaled conformal time $\eta$, defined by $a(\eta)
d\eta= (D-2)dt$, the solutions can be put in the simple form
\begin{eqnarray}
a(\eta)= A^{\frac{1}{D-2}} \left\{
\begin{array}{cr}
\left[\sin\left(\eta/2\right)\right]^{\alpha}&\kappa=1\\
\left(\eta\right/2)^{\alpha}&\kappa=0\\
\left[\sinh\left(\eta/2\right)\right]^{\alpha}&\kappa=-1
\end{array}\right.\, ,
\quad A=\frac{16\pi G_{N}\rho_0 R_0^D}{(D-1)(D-2)},  \alpha=\frac{2}{D-2}. \ \ \ \
\label{Dscalerd}
\end{eqnarray}
As can be seen from eq.~(\ref{Dscalerd}) the qualitative behavior of the solutions
does not depend strongly on $D$. In a (closed, open or flat) RD universe one always
has $R_{00}=G_{00}$, therefore $R_{CC}^{-2}=\frac{D-2}{2}\left(-\dot{H} +
\frac{D-4}{2}H^2 + \frac{D-2}{2}\frac{\kappa}{a^2}\right)$. The behaviour of
$S_{CEB}$ is easily derived from the explicit solution for  the scale factor and
$R_{CC}$. In the case D=4 it is shown in Fig.~\ref{fig1}.

A related case is when matter can be modelled by a conformal field theory (CFT).
Kutasov and Larsen \cite{Kutasov} pointed out that for weakly coupled CFT's in a
sphere of radius $R$, the free energy $F$, the entropy $S$ and the total energy $E$
can be expanded at weak coupling and large $x \equiv 2 \pi RT$ ,
\begin{eqnarray}
\label{Fexp}
 - FR &=& f(x) = \sum\limits_{n\geq 0} a_{D-2 n} x^{D-2 n} + \dots \,
\\
S &=& 2 \pi f'(x) \, ,\\
 ER &=& (x \partial_x - 1) f(x),
\end{eqnarray}
where the dots represent non-perturbative contributions.

We can explicitly check under which conditions the entropy of weakly coupled CFT's
obeys  the CEB, $S<S_{\rm CEB} = 4 B \sqrt{\pi} \sqrt{EV}l_P^{-(D-2)/2}$. In the
limit $TR \gg 1$ we find
\begin{eqnarray}
\frac{S^2}{S_{\rm CEB}^2} = \frac{\pi a_{D} D^2}{4 B^2 (D-1) \Omega_{D-1}} \left(2
\pi l_P T \right)^{D-2} \, .
\end{eqnarray}
Thus, CEB is obeyed provided that
\begin{eqnarray}\label{KD}
\left(\frac{ T}{M_P}\right)^{D-2}<  \frac{K(D)}{a_{D}},
\end{eqnarray}
where $K(D)$ is a $D$-dependent (but CFT independent) constant. We conclude that
CEB is obeyed as long as temperatures are below $M_P$ by a factor
$a_{D}^{-\frac{1}{D-2}}$ Since $a_{D}$ is proportional to the number $\N$ of
CFT-matter species, we obtain a bound on temperature which scales as
$\N^{-\frac{1}{D-2}}$ in Planck units.

We can also explicitly check under which conditions strongly coupled CFT's
possessing AdS duals as considered by Verlinde \cite{EV} obey the CEB. For such
CFT's,
\begin{eqnarray}
S=\frac{c}{12}\frac {V} {L^{D-1}}
\end{eqnarray}
\begin{eqnarray}
E=\frac{c}{12}\frac{D-1}{4\pi L}\left(1+\frac{L^2}{R^2}\right)\frac{V} {L^{D-1}}
\end{eqnarray}
\begin{eqnarray}
T=\frac{1}{4\pi L }\left(D +(D-2)\frac{L^2}{R^2}\right),
\end{eqnarray}
where $c$ is the central charge of the CFT and $L\sim 1/T$ is the AdS radius.

In this case, in the limit $R/L\sim TR\gg 1$ we find
\begin{eqnarray}
\frac{S^2}{S_{\rm CEB}^2}= \frac{1}{4(D-1) B^2}
\frac{c}{12}\left(\frac{l_P}{L}\right)^{D-2}
\end{eqnarray}
and thus CEB is obeyed for
\begin{eqnarray}\label{KD2}
\frac{1}{4(D-1) B^2} \frac{c}{12} \left(\frac{4 \pi T}{D M_P}\right)^{(D-2)}<1\, .
\end{eqnarray}
Since the central charge $c$ is proportional to the number of CFT fields $\N$, we
obtain a bound on temperature which, in Planck units, scales as
$\N^{-\frac{1}{D-2}}$, exactly as previously obtained for the weakly coupled case.

For the case $ER\sim a_{D}$ (which corresponds to $RT\sim 1$) the  validity of the
CEB guaranteed by a condition  similar to eq.~(\ref{KD}).

Finally, we would like to show that CEB holds also when $ER\sim 1$. In this case
$S_{CEB}\simeq 4B \sqrt{\pi} \sqrt{V/R} l_P^{-(D-2)/2}$ scales as
$\left(\frac{R}{l_P}\right)^{\frac{D-2}{2}}$. The appropriate setup for calculating
the entropy in this case is the microcanonical ensemble with the result $S\sim \log
a_D \sim \log \N$; thus $S<S_{CEB}$ is guaranteed for a macroscopic Universe as
long as
\begin{eqnarray}
\left(\frac{R}{l_P}\right)^{\frac{D-2}{2}}>\log \N\, .
\end{eqnarray}

In a quantum theory of gravity we expect the UV cut-off $\Lambda$ to be finite and
to represent an upper bound on $T$ (as in the example of superstring theory and its
Hagedorn temperature) and a lower bound on R (as in the minimal compactification
radius). Thus conditions (\ref{KD}), (\ref{KD2}) for the validity of CEB are
satisfied as long as $\left(\frac{\Lambda}{M_P}\right)^{D-2} < 1/\N$. A bound of
the same form was previously proposed in \cite{Bek3} and \cite{GSL}, and
independent arguments in support of bounds of this sort have also been  put forward
in \cite{shortest}.

\subsubsection{The inflationary Universe}

The inflationary universe is completely compatible with the CEB. To a certain
extent this is a not such an interesting case, because the CEB is comfortably
satisfied.

The entropy balance begins for the inflationary Universe after the end of inflation
when the energy of the background is converted to matter. This process is
historically called reheating and is associated with a large entropy production. In
the following we will assume that the reheating process is instantaneously and
complete. We will denote by the subscript $RH$ quantities at the instant of
reheating.

Since $\dot H$ is subleading in this case it follows from eq.~(\ref{rcch}) that
$R_{CC}\sim 1/H$. In this case the CEB and the HEB are similar,
\begin{equation}
S_{CEB}(t_{RH})= l_P^{-2} H(t_{RH}) a^3 (t_{RH}).
\end{equation}
Assuming that the energy has been completely converted into radiation, the energy
density of the radiation is $\rho(t_{RH})= T^4_{RH}$. From the $00$ Einstein
equation $l_P^{-2} H^2 = \rho$, thus
\begin{eqnarray}
S_{CEB}(t_{RH}) &=& T^4_{RH} \frac{1}{H(t_{RH})} a^3 (t_{RH}) \nonumber \\
&=& \frac{T_{RH}}{H(t_{RH})} T^3_{RH}a^3 (t_{RH}) \\
&=& \frac{T_{RH}}{H(t_{RH})} S(t_{RH}). \nonumber
\end{eqnarray}
Here we have used the expression for the radiation entropy $S(t_{RH})=T^3_{RH}a^3
(t_{RH})$. Since from the $00$ Einstein equation $\frac{T_{RH}}{H(t_{RH})}\sim
\sqrt{\frac{M_P}{H(t_{RH})}}$, and since we expect that the Hubble parameter at
reheat be substantially below the Planck temperature we conclude that the CEB is
comfortable satisfied.

\subsubsection{A Universe near a turning point}

Let us consider either a flat or closed universe with some perfect fluid in thermal
equilibrium and a constant equation of state $p=\gamma \rho~, 1>\gamma > -1$, and
with an additional small negative cosmological constant $\Lambda = - \lambda$. The
universe starts out expanding, reaches a maximal size, and then contracts towards a
singularity. In this case the matter entropy within a comoving volume is constant
in time. But near the point of maximal expansion the apparent horizon and the
Hubble length diverge causing violation of the HEB. However, for a fixed comoving
volume, $S_{CEB}\sim V R_{CC}^{-1}$, and, since $R_{CC}$ is never larger than some
maximal value, the CEB has a chance of doing better.

To see this explicitly let us consider a 4D example. In this case we obtain from
eq.~(\ref{rcch})
\begin{equation}
R_{CC}^{-2} = \frac{1}{3} {\rm Max} \left[ \frac{1}{2} \rho_0 (1-3\gamma)
a^{-3(1+\gamma)}- 2 \lambda~,~~ \frac{3}{2} (1 + \gamma) \rho_0
a^{-3(1+\gamma)}\right],
\end{equation}
independently of $\kappa$. The initial energy density is $\rho_0$ and $a$ is the
ratio of the scale factor to its initial value. Since the maximum is larger than
each of the expressions in the brackets
\begin{equation}
R_{CC}^{-2} \ge  \frac{1}{2} (1 + \gamma) \rho_0 a^{-3(1+\gamma)}.
\end{equation}
It follows that in a fixed comoving volume $S_{CEB}$ scales as $ \sim a^3
R_{CC}^{-1} \sim a^{3/2(1-\gamma)}$. Since $\gamma < 1$, this means that $S_{CEB}$
grows during the expansion, reaches a maximum at the turning point, and then starts
decreasing. If the initial conditions are fixed at at sufficiently early times when
curvature and cosmological constant are negligible, the CEB will be obeyed
initially provided energy density and curvature are less than Planckian. But then
the evolution of $S_{CEB}$ that we have found will guarantee that the bound is
satisfied at all times until Planckian density and curvature is reached in the
recollapsing phase. Thus the CEB will be satisfied throughout the classical
evolution of our Universe.

\subsubsection{A static Universe}

The simplest example of a non-singular cosmology is a static Einstein model in $D$
dimensions which was discussed in \cite{BFM}. This model requires positive
curvature, and two types of sources: cosmological constant and dust; we denote by
$\rho_{\Lambda}$ and $\rho_{\rm m}$ the energy densities associated with each of
the two components. To provide entropy we need an additional source, which we
choose to be radiation consisting of $\N$ species in thermal equilibrium at
temperature $T$. The energy density of the radiation is given by $\rho_{\rm r}=\N
T^D$, and the entropy density of the radiation is given by $s_{\rm r}=\N T^{D-1}$
(we ignore here numerical factors since we will be interested in scaling of
quantities). The total entropy of the system is given entirely by the entropy of
the radiation $S_{\rm r}=s_{\rm r} V$.

In term of these sources, Einstein's equations can be written in the following way:
\begin{eqnarray}\label{ein1}
H^2 + \frac{1}{a^2}&=& \frac{16 \pi G_{\rm N}}{(D-2)(D-1)}\rho_{\rm tot}= \frac{16
\pi G_{\rm N}}{(D-2)(D-1)}\left(\rho_{\Lambda}+\rho_{\rm m}+
\rho_{\rm r}\right)\\
\label{ein2} \dot{H} - \frac{1}{a^2}&=& -\frac{8\pi G_{\rm
N}}{(D-2)}\left(\rho_{\rm tot} + p_{\rm tot}\right) \nonumber \\ &=& -\frac{8\pi
G_{\rm N}}{(D-2)(D-1)} \left[D\rho_{\rm r}+(D-1)\rho_{\rm m}\right]\, ,
\end{eqnarray}
where we have used in eq.~(\ref{ein2}) the equations of state relating pressure to
energy density: $p_{\Lambda}=-\rho_{\Lambda}$, $p_{\rm m}=0$, and $(D-1)p_{\rm
r}=\rho_{\rm r}$.

For given $\rho_{\rm m}$ and $\rho_{\rm r}$, one can choose $\rho_{\Lambda}$ and
the scale factor $a$ such that $H$ and $\dot{H}$ vanish in eqs.~(\ref{ein1}) and
(\ref{ein2}), and thus obtains a static solution. In particular, the condition
given by eq.~(\ref{ein2}) determines the scale factor in terms of $\rho_{\rm
  m}$ and $\rho_{\rm r}$,
\begin{equation}\label{e2vanish}
a^2=\frac{(D-2)(D-1)}{8\pi G_{\rm N}}\frac{1}{D\rho_{\rm r} + (D-1)\rho_{\rm m}}\,
.
\end{equation}
Since both $H$ and $\dot H$ vanish identically, $R_{\rm CC}$ is determined solely
by the scale factor $a$ given in eq.~(\ref{e2vanish}), as discussed previously.

We now wish to determine under which conditions (if any) some violations of CEB may
occur in this model. Recall that according to eq.~(\ref{cebd1}) the CEB bounds the
total entropy of a region contained in a comoving volume $V$ by $S_{\rm CEB}=\alpha
(D-1)\frac{V}{G_{\rm N} R_{\rm CC}}$,  and that in the static case under
consideration $R_{\rm CC}=2 a/(D-2)$. The square of the ratio of $S_{\rm CEB}$ and
the entropy of the system $S_{\rm r}$, is given by
 \begin{eqnarray}
 \label{Soverceb}
\left(\frac{S_{\rm CEB}}{S_{\rm r}}\right)^2 &=& \left( \frac{\alpha(D-1)} {s_{\rm
r} R_{\rm CC}
G_{\rm N}}\right)^2 \nonumber \\
&=& \left[ 2\pi \alpha^2 (D-1)(D-2)\right] \left[{D + (D-1){\rho_{\rm m}\over
\rho_{\rm r}}}\right] \left[\frac{1}{\N} \left(\frac{M_P}{T}\right)^{D-2}\right]. \
\ \ \ \ \ \
\end{eqnarray}
Since the second factor in expression (\ref{Soverceb}) is larger than unity if
$\rho_{\rm m}$ and $\rho_{\rm r}$ are positive, and neglecting the overall
prefactor which is independent of the sources in the model, we conclude that the
CEB is valid provided that
\begin{equation}\label{subP}
 \N\left({T\over M_{\rm P}}\right)^{D-2}\le 1.
\end{equation}
This is the same condition discussed above which should be interpreted as a
requirement that temperatures are sub-Planckian, in the case of many number of
species $\N$.

Our conclusion is that as long as the temperature of radiation stays well below
Planckian, CEB is upheld. The fact that the model is gravitationally unstable to
matter perturbations does not seem to be particularly relevant to the issue of
validity of the CEB.

\subsubsection{Bekenstein's non-singular Universe}

A time-dependent non-singular cosmological model was found years ago by Bekenstein
\cite{bmodel} (see also \cite{mayo}). This is a 4D Friedman-Robertson-Walker
universe which is conformal to the closed Einstein Universe. It contains dust,
consisting of $N$ particles of mass $\mu$ ($N$ is constant and $\mu$ is positive),
coupled to a classical conformal massless scalar field $\psi$, and $\N$ species of
radiation in thermal equilibrium. The action for the dust-$\psi$ system is given by
\begin{eqnarray}
{\cal S}=-\frac{1}{2}\int\sqrt{-g}\left[\left(\nabla\psi\right)^2 + \frac{1}{6}
\psi^2 R\right]d^4 x - \int\left(\mu + f\psi\right)d\tau.
\end{eqnarray}
It includes in addition to the usual action for free point particles of rest mass
$\mu$, a dust-scalar field interaction whose strength is determined by the coupling
$f$. Accordingly, we may define the effective mass of the dust particles: $\mu_{\rm
eff}=\mu+f \psi$.

The total energy density and pressure in Bekenstein's Universe are given by
\begin{eqnarray}\label{rototmay}
\rho_{\rm tot}=\rho_{\rm r}+\rho_{\psi}+\rho_{\rm m},\quad p_{\rm tot}&=&p_{\rm
r}+p_\psi+p_{\rm m} ,
\end{eqnarray}
where $\{\rho_{\rm r},p_{\rm r}\}$, $\{\rho_{\psi},p_\psi\}$, and $\{\rho_{\rm
m},p_{\rm m}\}$ are the energy densities and pressures associated with the
radiation, scalar field and dust respectively. They depend on the scale factor in
the following way
\begin{eqnarray}
\label{equations}
\rho_{\rm r}&=&{\cal C} \N a^{-4}=\N T^4, \nonumber \\
 \rho_\psi &=& {1\over 2}f^2N^2 a^{-4},  \\
 \rho_{\rm m} &=&N\mu_{\rm eff}
a^{-3}=N\mu a^{-3} -2\rho_\psi, \nonumber
\end{eqnarray}
and their equations of state $\gamma_{\rm r}=p_{\rm r}/\rho_{\rm r}$,
$\gamma_{\psi}=p_\psi/\rho_{\psi}$,  $\gamma_{\rm m}=p_{\rm m}/\rho_{\rm m}$ are
the following
\begin{eqnarray}
\label{eoss}
\gamma_{\rm r}&=&1/3, \nonumber \\
 \gamma_\psi &=& -1/3,  \\
 \gamma_{\rm m} &=&0. \nonumber
\end{eqnarray}
The dependence of $\psi$ on $a$  $\psi=-f N a^{-1}$, yields $\mu_{\rm eff}=\mu-f^2
N a^{-1}$. ${\cal C}$ is an integration constant and the only source of entropy is
the radiation whose entropy density is given by $s_{\rm r}=\N T^3$.

The solution for the scale factor $a$ is given in terms of the conformal time
$\eta$ by
\begin{equation}
a(\eta)=a_0(1+B \sin\eta).
\end{equation}
We assume that $a_0$, the mean value of the scale factor, is macroscopic, so it is
large in our Planck units. If $B=0$ the solution describes a static universe very
similar to the closed Einstein Universe discussed previously. For $0<B<1$ the
solution describes a ``bouncing universe": the universe bounces off at $\eta=3
\pi/2$ when the scale factor is minimal $a=a_{\rm min}=a_0(1-B)$, expands until it
turns over at $\eta=5\pi/2$ when its scale factor is maximal $a=a_{\rm
max}=a_0(1+B)$, and continues to oscillate without ever reaching a singularity. The
equations of motion require that the energy densities of the sources obey the
following equalities at all times \cite{bmodel}:
\begin{equation}
2 {a\over a_0} \left({\rho_\psi-\rho_{\rm r}\over2\rho_\psi+ \rho_{\rm m}}
\right)=1-B^2={a_{\rm min} a_{\rm max}\over {a_0}^2}.
 \label{equality}
\end{equation}
Since $2\rho_\psi+\rho_{\rm m}=N\mu a^{-3}>0$, $\rho_{\rm r}>0$, and $B^2<1$, it
follows that a necessary condition for a bounce is that $\rho_{\rm r}<\rho_\psi $.
This implies that the total pressure $\frac{1}{3} (\rho_{\rm r}-\rho_{\psi})$ is
always negative. Moreover, eq.~(\ref{equality}) for $a=a_{\rm min}$ implies that
$\rho_{\rm m}\leq-2\rho_{\rm r}<0$ there. But then, the conclusion must be that in
order to avoid a singularity, $\mu_{\rm eff}<0$ at least at the bounce. It is
possible, however, to find a range of initial conditions and parameters such that
$\mu_{\rm eff}$ is positive near the turnover.

The result that $\rho_{\rm r}$ and $\rho_\psi$ are manifestly positive definite,
but $\rho_{\rm m}$ can (and in fact must) be negative some of the time, suggest
that it might be possible to parametrically decrease $\rho_{\rm tot}$ by lowering
$\mu_{\rm eff}$ (making it large and negative) by increasing the coupling strength
$f$, so that the amounts of radiation and entropy are kept constant. As it turns
out this is exactly the case in which the CEB can be potentially violated. Using
Einstein's equations to express $R_{\rm CC}$ in terms of the total energy density
and pressure, we find the ratio $\left(S_{\rm CEB}/S_{\rm r}\right)^2$:
\begin{eqnarray}
 \label{rat1}
 \left({S_{\rm CEB}\over S_{\rm r}}\right)^2 \sim {G_{\rm N}}^{-2}
 \left(\frac{\rho_{\rm r}}{\cal N}\right)^{-3/2}
 \frac{1}{{\cal N}^2} G_{\rm N}\,
 {\rm Max} \left[
 {\rho_{\rm tot}\over 3}-p_{\rm tot} ,\rho_{\rm tot}+p_{\rm tot}
 \right].
\end{eqnarray}
A system for which the ratio above is smaller than one would violate the CEB.
Recalling that the maximum on the r.h.s. of (\ref{rat1}) is always larger than the
mean of the two entries and rearranging we find
\begin{equation}
\left({S_{\rm CEB}\over S_{\rm r}}\right)^2 \geq
\left[\frac{1}{\N}\frac{M_P^2}{T^2}\right]{\rho_{\rm tot}\over \rho_{\rm r}}.
\label{ratio}
\end{equation}
Since we assume that the model is sub-Planckian, namely that the first factor is
larger than one as in eq.~(\ref{subP}), the only way in which CEB could be violated
is if somehow the second factor was parametrically small. As discussed above, it
does seem that the second term $\rho_{\rm tot}/\rho_{\rm r}$ can be made
arbitrarily small by decreasing $\rho_{\rm tot}$ while keeping $\rho_{\rm r}$
constant. Consequently, it is apparently possible to make the ratio $S_{\rm
  CEB}/ S_{\rm r}$ smaller than one and obtain a CEB violating
cosmology. But this can be achieved only if the effective mass of the dust
particles is negative (and large) as can be seen from eq.~(\ref{rototmay}).

Violations of the CEB (and as a matter of fact, of any other entropy bound) go hand
in hand with large negative energy densities in the dust sector. In the model under
discussion, this manifests itself in the form of dust particles with highly
negative effective masses. Occurrence of such negative energy density would most
probably render the model unstable.  We argue that any analysis of entropy bounds
should be performed for stable models. This is particularly relevant for the CEB,
whose definition involves explicitly the largest scale at which stable BH's could
be formed. However, the instability does not necessarily lead to violations of the
CEB as in the previous case. To support this argument we have outlined possible
instabilities in the dust scalar field system when the dust particles mass is
negative \cite{BFM}.

\subsubsection{The pre-big-bang scenario}

Veneziano was the first to study entropy bounds in the context of the PBB scenario
\cite{HEB}.  It has been argued \cite{GV,BDV} that a form of stochastic PBB is a
generic consequence of  natural initial conditions corresponding to generic
gravitational and dilatonic waves superimposed on the perturbative vacuum of
critical superstring theory. In the Einstein-frame metric this can be seen as a
chaotic gravitational collapse leading to the formation of BH's of different sizes.
For a string frame observer inside each BH this is viewed as a PBB inflationary
cosmology.  The duration of the inflationary phase is controlled by the size of the
BH \cite{GV,BDV}, so from this point of view the observable Universe should be
identified with the region of space that was originally inside a sufficiently large
BH.

In \cite{HEB} Veneziano studied a 4D PBB model and followed the evolution of
several contributions to the entropy. At time $t=t_i$, corresponding to the first
appearance of a horizon, he used the Bekenstein--Hawking formula to evaluate that
the entropy in the collapsed region $S_{coll}$. Then he used the fact \cite{BDV}
that the initial size of the BH horizon determines the initial value of the Hubble
parameter and found that
\begin{equation}
 S_{coll} \sim (R_{in}/l_{P,in})^2 \sim (H_{in} l_{P,in})^{-2}  = S_{HEB}.
\label{initialS}
\end{equation}
Thus, initially the entropy is as large as allowed by the HEB (without
fine-tuning). Here it was implicitly assumed the initial string coupling is small.

After a short transient phase, dilaton-driven inflation (DDI) should follow
\cite{GV,BDV} and last until $t_s$, the time at which a string-scale curvature is
reached. We expect this classical process not to generate further entropy. During
DDI $S_{HEB}$ remains constant and the bound continues to be saturated. This
follows from the  ``conservation law" of string cosmology \cite{PBBrev}
\begin{equation}
\partial_t \left(e^{-\phi} \sqrt{g} H \right) = 0,
\end{equation}
hence
\begin{equation}
\partial_t \left((\sqrt{g} H^3)~~( e^{-\phi} H^{-2}) \right) =
\partial_t \left( n_H S^H \right) = 0~.
\label{conservation}
\end{equation}
Veneziano suggested the following interpretation: At the beginning of the DDI phase
the whole entropy is in a single Hubble volume. As DDI proceeds, the same total
amount of entropy becomes equally shared between very many Hubble volumes until,
eventually, each one of them contributes a small number.

While the coupling is still small $S_{HEB}$ cannot decrease,
\begin{equation}
\partial_t(e^{-\phi} \sqrt{g} H) \ge 0.
\label{HEBgrowth}
\end{equation}
It follows that
\begin{equation}
(\dot{\phi} - 3H) \le \dot{H}/H \; . \label{HEBgrowth1}
\end{equation}
Veneziano noticed that this constraint may be important. As $\alpha'$ corrections
intervene to stop the growth of $H$, the entropy bound forces $\dot{\phi} - 3H$ to
decrease and eventually to change sign if $H$ stops growing. But this is just what
is needed to convert the DDI solution into the FRW solution \cite{PBBrev}.

If the initial conditions are such that the string coupling becomes strong while
the curvature is still small then Veneziano argued \cite{HEB} that the HEB forces a
non-singular PBB cosmology as well. This time the entropy production by the
squeezing of quantum fluctuations is the important factor. This will be discussed
further when we discuss the generalized second law.

\subsection{Conditions for the validity of the CEB in cosmology}

We may summarize the lessons of the previous examples by imposing conditions on
sources in a generic cosmological setting such that the CEB is obeyed.

We consider a cosmic fluid consisting of radiation, an optional cosmological
constant, and additional unspecified classical dynamical sources which do not
include any contributions from the cosmological constant or radiation. For
simplicity we assume that the additional sources have negligible entropy. This is
the most conservative assumption: if some of the additional sources have
substantial entropy our conclusions can be strengthened. We use the previous
notations for the total, cosmological, and radiation energy densities, $\rho_{\rm
tot}$, $\rho_{\Lambda}$ and $\rho_{\rm r}$ respectively, and denote by $\rho^*$ the
combined energy density of the additional sources. Thus
\begin{equation}
\rho_{\rm tot}=\rho_{\rm r}+\rho_{\Lambda}+\rho^*.
\end{equation}
We use the same notation for the relative pressures, and for the equation of state
$\gamma^*\equiv\rho^*/p^*$, which may be time-dependent.

In term of these sources, the causal connection scale can be written as
\begin{eqnarray}\label{Rccrho}
&& R_{\rm CC}^{-2}=\frac{4\pi G_{\rm N}}{D-1} {\rm Max} \Biggl\{D\rho_{\Lambda} +
\biggl[1 - (D-1)\gamma^*\biggr]\rho^*\, , \nonumber \\ && (D-4)\rho_{\Lambda} +
\biggl[(2D-5) + (D-1)\gamma^*\biggr]\rho^* + 2(D-2)\rho_{\rm r}\Biggr\}.
\end{eqnarray}

We may now express the ratio of $(S_{\rm CEB}/S_{\rm r})^2$, neglecting as usual
prefactors of order one
\begin{eqnarray}
&& \left(\frac{S_{\rm CEB}}{S_{\rm r}}\right)^2 \sim \frac{1}{\N}\left(\frac{M_{\rm
P}}{T}\right)^{D-2}\nonumber  {\rm Max}\Biggl\{ D\frac{\rho_{\Lambda}}{\rho_{\rm
r}}+ \biggl[ 1 - (D-1)\gamma^*\biggr]\frac{\rho^*}{\rho_{\rm r}}, \nonumber \\ &&
(D-4)\frac{\rho_{\Lambda}}{\rho_{\rm r}}+\biggl[ (2D-5) +
(D-1)\gamma^*\biggr]\frac{\rho^*}{\rho_{\rm r}} + 2(D-2)\Biggr\} .\label{ratiogen}
\end{eqnarray}
Any CEB violations requires that this ratio be parametrically smaller than one.
Notice that the first factor is larger than one by our requirement that the
radiation energy density be sub-Planckian. Thus the only remaining possibility for
violating CEB is that the second factor be parametrically smaller than unity. As we
show below, this can occur only if at least one of the additional sources has
negative energy density.

The r.h.s. of (\ref{ratiogen}) is larger than the average of the two entries, so
that
\begin{equation}
\label{newrat} \left(\frac{S_{\rm CEB}}{S_{\rm r}}\right)^2 \geq
{1\over\N}\left({M_{\rm P}\over T}\right)^{D-2} (D-2) {\rho_{\rm tot}\over\rho_{\rm
r}}.
\end{equation}
Therefore, since $\rho_{\rm tot}>0$, a necessary condition for this expression to
be smaller than unity is that $\rho_{\rm
  tot}\ll\rho_{\rm r}$, which we may reexpress as
\begin{eqnarray}\label{condL}
\frac{\rho_{\Lambda}}{\rho_{\rm r}}\sim -\left(1+\frac{\rho^*}{\rho_{\rm
r}}\right).
\end{eqnarray}
This is not a sufficient condition since the equations of motion could dictate, for
example, that the first factor on the r.h.s. of eq.~(\ref{newrat}) could be
parametrically larger than unity at the same time. By substituting condition
(\ref{condL}) into eq.~(\ref{ratiogen}), we obtain
\begin{eqnarray}
&& \left(\frac{S_{\rm CEB}}{S_{\rm r}}\right)^2 \sim \frac{1}{\N}\left(\frac{M_{\rm
P}}{T}\right)^{D-2} \times \nonumber \\ && {\rm
Max}\left\{-\biggl[(D-1)(1+\gamma^*)\frac{\rho^*} {\rho_{\rm r}} +
D\biggr]\,,(D-1)(1+\gamma^*)\frac{\rho^*} {\rho_{\rm r}} + D\right\}
.\label{ratnongen}
\end{eqnarray}
Therefore, an additional necessary  condition for $S_{CEB}/S_{\rm r}$ to be smaller
than one is that
\begin{eqnarray}\label{cond*}
(1 + \gamma^*)\rho^*\simeq-\frac{D}{(D-1)}\rho_{\rm r}\, .
\end{eqnarray}
Condition (\ref{cond*}) can be satisfied in two ways:

(i) $1+\gamma^*>0$ and $\rho^*<0$. This obviously requires that at least one of the
sources has negative energy density. In this case (barring pathologies) the
magnitude of $\rho^*$ is comparable to that of $\rho_{\rm r}$.

(ii) $1+\gamma^*<0$ and $\rho^*>0$. However, for classical dynamical sources,  this
typically clashes with causality which requires that the pressure and energy
density of each of the additional dynamical sources obey $|p_i|<|\rho_i|$; hence if
all $\rho_i>0$ then necessarily \hbox{$\gamma^*=\left(\sum p_i\right)/\left(\sum
\rho_i\right)>-1$}.

Consequently, condition (\ref{cond*}) cannot be satisfied if all of the dynamical
sources have positive energy densities and equations of state $|\gamma_i|\le 1$.
Bekenstein's Universe discussed previously fits well within our framework: the
total energy density is positive, but the overall contribution to $\rho_{\rm tot}$
of all the sources, excluding radiation (since the cosmological constant vanishes
in this case), is negative and almost cancels the contribution of radiation,
leaving a small positive $\rho_{\rm tot}$.

To summarize, if all dynamical sources (different from the cosmological constant)
have positive energy densities $\rho_i > 0$ and have causal equations of state
($|\gamma_i|\le 1$), and if radiation temperatures are sub-Planckian, CEB is
upheld.

\subsection{The CEB and the singularity theorems}

The CEB (and entropy bounds in general) refines the classic singularity theorems.
It is satisfied by cosmologies for which the singularity theorems are not
applicable because some of the energy conditions are violated, but do not seem to
be problematic in any of their properties. Conversely, it indicates possible
problems when the singularity theorems seem perfectly valid.

In general, the total energy-momentum tensor of a closed ``bouncing" universe
violates the SEC, but it can obey the CEB. In order to see this explicitly  let us
consider the ``bounce'' condition, i.e.~ $H=0$, $\dot{H}>0$ for a closed Universe;
by using the Einstein equations (\ref{ein1}-\ref{ein2}), we can express this
condition in terms of the sources as follows:
\begin{eqnarray}\label{bounce}
\rho_{\rm tot}>0,\quad (D-3)\rho_{\rm tot}+(D-1)p_{\rm tot}<0.
\end{eqnarray}
The second of these conditions is (in $D=4$) precisely the condition for violation
of the SEC.  In terms of $\rho_{\rm r}$, $\rho_{\Lambda}$ and $\rho^*$ this reads
\begin{equation}
\label{condbounce} 2 \rho_{\Lambda} - (D-2)\rho_{\rm r} - \biggl[(D-3) +
(D-1)\gamma^*\biggr]\rho^*>0\, .
\end{equation}
In comparison, a necessary condition that the CEB is violated can be obtained from
eqs.(\ref{condL}) and (\ref{cond*}),
\begin{equation}
\label{3.10} 2 \rho_{\Lambda} - (D-2)\rho_{\rm r} - \biggl[(D-3) +
(D-1)\gamma^*\biggr]\rho^*\sim 0\, ,
\end{equation}
where the l.h.s of (\ref{3.10}) can be either positive or negative. So we find that
there is a range of parameters for which the CEB can be obeyed in  some bouncing
cosmologies but not in others.

In a spatially flat universe ($\kappa=0$), the conditions for a bounce are slightly
different: $\rho_{\rm tot}=0$ and $\rho_{\rm tot}+p_{\rm tot}<0$. At the bounce
these conditions imply violation of the Null Energy Condition (NEC). As discussed
previously, classical sources are not expected to violate the NEC, but effective
quantum sources  (such as Hawking radiation) are known to violate the NEC. In terms
of $\rho_{\rm r}$, $\rho_{\Lambda}$ and $\rho^*$ the condition for a bounce reads
\begin{equation}
\left(1+\frac{1}{D-1}\right) \rho_{\rm r} + (1+\gamma^*)\rho^*>0.
\end{equation}
In comparison, a necessary condition that the CEB is violated can be obtained from
eq.~(\ref{cond*}),
\begin{equation}
\label{3.12} \left(1+\frac{1}{D-1}\right) \rho_{\rm r} + (1+\gamma^*)\rho^*\sim 0\,
,
\end{equation}
where the l.h.s of (\ref{3.12}) can be either positive or negative. So, again, we
find that there is a range of parameters for which the CEB can be obeyed in  some
spatially flat bouncing cosmologies but not in others.

The CEB appears to be a more reliable criterion than energy conditions when trying
to decide whether a certain cosmology is reasonable: taking again the closed
deSitter Universe as an example, we can add a small amount of radiation to it, and
still have a bouncing model if $\rho_{\Lambda}$ is the dominant source, and SEC
will not be obeyed (see eq.~(\ref{condbounce})). Nevertheless, the general
discussion in this section shows that in this case the CEB is not violated as long
as radiation temperatures remain subPlanckian, despite the presence of a bounce.
This happens, in part, because the CEB  is able to discriminate better between
dynamical and non-dynamical sources (such as the cosmological constant), and
imposes constraints that involve the former ones only, such as eq.~(\ref{cond*}).

We have reached the following conclusions by studying the validity of the CEB for
non-singular cosmologies:
 \begin{enumerate}

\item Violation of the CEB necessarily requires either high temperatures ${\cal N}
\left(\frac{T}{M_{\rm P}}\right)^{D-2} \geq 1$, or dynamical sources that have
negative energy densities with a large magnitude, or sources with acausal equation
of state. Of course, neither of the above is sufficient to guarantee violations of
the CEB.

\item Classical sources of this type are suspect of being unphysical or unstable,
but each source has to be checked on a case by case basis. In the examples that we
have discussed the sources were indeed found to be unstable or are strongly
suspected to be so.

\item Sources with large negative energy density could allow, in principle, to
increase the entropy within a given volume, while keeping its boundary area and the
total energy constant. This would lead to violation of all known entropy bounds,
and of any entropy bound which depends in a continuous way on the total energy or
on the linear size of the system.

\item The CEB is more discriminating than singularity theorems. In the examples we
have considered it allows non-singular cosmologies for which singularity theorems
cannot be applied, but does not allow them if they are associated with specific
dynamical problems.

\end{enumerate}

\subsection{Comparison of the CEB to other entropy bounds}

Finally, we compare our CEB to other bounds, in particular to Bekenstein's and
Bousso's. For systems of limited gravity whose size exceeds their Schwarzschild
radius: $R > R_g$, Bekenstein's bound is given by $S<S_{BEB} = l_P^{-2} R~R_g$, and
Bousso's procedure results in the holography bound, $S < S_{HOL} = l_P^{-2} R^2$,
but since $R > R_g$, $S_{BEB}<S_{HOL}$, and therefore Bousso's bound is less
stringent than Bekenstein's. Consider now the CEB applied to the region of size $R$
containing an isolated system. Expressing CEB in the form (\ref{CCB}) one
immediately obtains: $ S_{CEB} = l_P^{-1} R^{3/2} E^{1/2} \hbar^{-1/2} =
(S_{HOL}~~S_{BEB})^{1/2}\; , $ implying $ S_{BEB}\le S_{CEB} \le S_{HOL}. $ We
conclude that for isolated systems of limited self-gravity the Bekenstein bound is
the tightest, followed by our CEB and, finally, by Bousso's holographic bound.
Similar scaling properties for the HEB were discussed in \cite{HEB}.

For regions of space that contain so much energy that the corresponding
gravitational radius $R_g$ exceeds $R$, Bekenstein's bound is the weakest, while
the naive holography bound is the strongest (but very often wrong). Bousso's
proposal uses the apparent horizon $R_{AH}$ while CEB uses $R_{CC}$. For
homogeneous cosmologies, $R_{CC}<R_{AH}$, since $R_{CC}^{-2}$, according to
(\ref{RCCk}), is always larger than the average of the two terms appearing on its
r.h.s., which is precisely $R_{AH}^{-2} = H^2+\kappa/a^2$. Since, for a fixed
volume, the bounds scale like $R_{AH}^{-1}$ or $R_{CC}^{-1}$, we immediately find
that CEB is generally more generous. An important difference between our proposal
and Bousso's covariant holographic bound \cite{boussorev} that scales as $S/A$ is
that there the entropy $S$ is a flux through light-like hypersurfaces. A detailed
comparison with Bousso's proposal is therefore more subtle because of his use of
the apparent horizon area to bound entropy on light sheets. This can be converted
into a bound on the entropy of the space-like region only in special cases.

E. Verlinde \cite{EV} argued that the radiation in a closed, radiation dominated
Universe can be modelled by a CFT, and that its entropy can be evaluated using a
generalized Cardy formula. After an appropriate modification of Verlinde's bound
which evades the criticism about its validity for weakly coupled CFT's the new
bound is exactly equivalent to CEB within the CFT framework.

\section{The Generalized second law and the Causal Entropy Bound}
\label{sec:4}

\subsubsection{The Generalized second law in Cosmology}

There seems to be a close relationship between entropy bounds and the GSL. We have
proposed a concrete classical and quantum mechanical form of the GSL in cosmology
\cite{GSL}, which is valid also in situations far from thermal equilibrium. We
discuss various entropy sources, such as thermal, ``geometric" and ``quantum"
entropy, apply GSL to study cosmological solutions, and show that it is compatible
with entropy bounds. GSL allows a more detailed description of how, and if,
cosmological singularities are evaded. The proposed GSL is different from GSL for
BH's \cite{gslbh}, but the idea that in addition to normal entropy other sources of
entropy have to be included has some similarities. We will discuss here only 4D
models. Obviously it should be possible to generalize our analysis to higher
dimensions in a straightforward manner along the lines of the generalizations of
the CEB to higher dimensions.

The starting point of our classical discussion is the definition of the total
entropy of a domain containing more than one cosmological horizon \cite{HEB}. We
have already introduced the number of cosmological horizons within a given comoving
volume $V=a(t)^3$. It is simply the total volume divided by the volume of a single
horizon, $n_H={a(t)^3}/{ |H(t)|^{-3}}$. As usual we will ignore numerical factors
of order unity. Here we use units in which $c=1, G_{N}=1/16\pi, \hbar=1$ and
discuss only flat, homogeneous, and isotropic cosmologies.  If the entropy within a
given horizon is $S^H$, then the total entropy is given by $ S=n_H S^H$. Classical
GSL requires that the cosmological evolution, even when far from thermal
equilibrium, must obey $
 dS \ge 0,
$ in addition to Einstein's equations. In particular,
\begin{equation}
 n_H \partial_t S^H+ \partial_t n_H S^H \ge 0.
\label{myeq}
\end{equation}

In general, there could be many sources and types of entropy, and the total entropy
is the sum  of their contributions.  If, in some epoch, a single type of entropy
makes a dominant contribution to $S^H$, for example, of the form $S^H= |H|^\alpha$,
$\alpha$ being a constant characterizing the type of entropy source, and therefore
$S=(a|H|)^3 |H|^\alpha$, eq.~(\ref{myeq}) becomes an explicit inequality,
\begin{equation}
 3H+(3+\alpha)\frac{{\dot H}}{H} \ge 0,
 \label{ec}
\end{equation}
which can be translated into energy conditions constraining the energy density
$\rho$, and the pressure $p$ of (effective) sources. Using the FRW equations,
\begin{eqnarray}
 H^2 &=&\ \frac{1}{6}\ \rho \nonumber \\
 {\dot H} &=& -\frac{1}{4} (\rho+p) \label{frw} \\
 {\dot \rho} &+& 3H (\rho+p)=0, \nonumber
\end{eqnarray}
and assuming $\alpha>-3$ (which we will see later is a reasonable assumption ) and
of course $\rho>0$, we obtain
\begin{eqnarray}
 \frac{p}{\rho}&\le&\frac{2}{3+\alpha }-1
 \hspace{.25in} \hbox{for} \hspace{.25in}
 H>0,
 \label{adeqp}\\
 \frac{p}{\rho}&\ge&\frac{2}{3+\alpha }-1
 \hspace{.25in} \hbox{for} \hspace{.25in} H<0.
 \label{adeqn}
\end{eqnarray}
Adiabatic evolution occurs when the inequalities in eqs.(\ref{adeqp},\ref{adeqn})
are saturated.

A few remarks about the allowed range of values of $\alpha$ are in order. First,
the usual adiabatic expansion of a radiation dominated universe with $p/\rho=1/3$
corresponds to $\alpha=-3/2$. Adiabatic evolution with $p/\rho<-1$, for which the
null energy condition is violated would require a source for which $\alpha<-3$.
This is problematic since it does not  allow a flat space limit of vanishing $H$
with finite entropy. The existence of an entropy source with $\alpha$ in the range
$\alpha<-2$ does not allow a finite $\partial_t S$ in the flat space limit and is
therefore suspected of being unphysical.  Finally, the equation of state
 $p=-\rho$ (deSitter inflation),  cannot be
described as adiabatic evolution for any finite $\alpha$.

Let us discuss in more detail three specific examples. First, as already noted, we
have verified  that thermal entropy during radiation dominated (RD) evolution can
be described without difficulties, as expected. In this case,
$\alpha=-\frac{3}{2}$, reproduces the well known adiabatic expansion, but also
allows entropy production. The present era of matter domination requires a more
complicated description since in this case one source provides the entropy, and
another source the energy.

The second case is that of geometric entropy $S_g$, whose source is the existence
of a cosmological horizon \cite{gibbons,Srednicki:1993im}. The concept of geometric
entropy is closely related to the holographic principle and to entanglement entropy
(see below). For a system with a cosmological horizon $S^H_g$ is given by (ignoring
numerical factors of order unity)
\begin{equation}
 S_{g}^{H}=|H|^{-2} G_N^{-1}.
 \label{gentropy}
\end{equation}
The equation of state corresponding to adiabatic evolution with dominant $S_g$, is
obtained by substituting $\alpha=-2$ into eqs.(\ref{adeqp},\ref{adeqn}), leading to
$p/\rho=1$ for positive and negative $H$. This equation of state is simply that of
a free massless scalar field, also recognized as  the two vacuum branches of PBB
string cosmology \cite{PBBrev} in the Einstein frame.  In \cite{HEB} this was found
for the $(+)$ branch in the string frame as an ``empirical" observation. In
general, for the case of dominant geometric entropy, GSL requires, for positive
$H$, $
 p \le \rho,
$ hence deSitter inflation  is definitely allowed. For negative $H$, GSL requires $
 \rho \le p,
$ and therefore forbids, for example, a time reversed history of our universe or a
contracting deSitter universe with a negative constant $H$ (unless some additional
entropy sources appear).

The third case is that of quantum entropy $S_{q}$, associated with quantum
fluctuations. This form of entropy was discussed in \cite{BMP,gg}. Specific quantum
entropy for a single physical degree of freedom is approximately given by (again,
ignoring numerical factors of order unity)
\begin{equation}
 s_{q}= \int d^3k \ln n_k,
 \label{quantums}
\end{equation}
where $n_k \gg 1$ are occupation numbers of quantum modes. Quantum entropy  is
large for highly excited quantum states, such as the squeezed states obtained by
amplification of quantum fluctuations during inflation. Quantum entropy does not
seem to be expressible in general as $S^H_{q}= |H|^{\alpha}$, since occupation
numbers depend on the whole history of the evolution. We will discuss this form of
entropy in more detail later, when the quantum version of GSL is proposed.

Geometric entropy is related to the existence of a horizon or more generally to the
existence of a causal boundary. From my current perspective the geometric entropy
corresponds to entanglement entropy of fluctuations whose wavelength is shorter
than the horizon while ``quantum" entropy is probably related to entanglement
entropy of fluctuations whose wavelength is larger than the horizon (see below).

We would like to show  that it is possible to formally define a temperature, and
that the definition is compatible with the a generalized form of the first law of
thermodynamics (see also \cite{Jacobson}). Recall that the first law for a closed
system states that $
 T dS = dE+pdV = (\rho+p)dV +V d\rho.
$ Let us now consider the case of single entropy source and formally define a
temperature $T$, $
 T^{-1}=\left( \frac{\partial S}{\partial E}\right)_V
 =\frac{\partial s}{\partial \rho },
$ since $E=\rho V$ and $S=sV$.  Using eqs.(\ref{frw}), and $s=|H|^{\alpha+3}$, we
obtain $
 \frac{\partial s}{\partial \rho}=
 \frac{\alpha+3}{12} |H|^{\alpha+1},
$ and therefore
\begin{equation}
 T=\frac{12}{\alpha+3} |H|^{-\alpha-1}.
 \label{gent}
\end{equation}
To ensure  positive temperatures $ \alpha > -3, $ a condition which we have already
encountered. Additionally,
 for $\alpha>-1$,  $T$ diverges in the flat space
 limit, and therefore such a source is
 suspect of being unphysical, leading to the conclusion that the
 physical range of $\alpha$ is $-2\le \alpha\le-1$.
A compatibility check requires $
 T^{-1}=
 \frac{ \partial s}{\partial t} / \frac{\partial \rho}{\partial t}
$, which indeed yields a result in agreement with (\ref{gent}). Yet another
thermodynamic relation $ p/T=\left( \frac{\partial S}{\partial V}\right)_E$, leads
to $p=sT-\rho$ and therefore to $
 p/\rho =\frac{2}{\alpha+3}-1
$ for adiabatic evolution, in complete agreement with
eqs.(\ref{adeqp},\ref{adeqn}). For $\alpha=-2$, eq.~(\ref{gent}) implies $T_g=|H|$,
in agreement with \cite{gibbons},  and  for ordinary thermal entropy $\alpha=-3/2$
reproduces the known result, $T=|H|^{1/2}$.

Is GSL compatible with entropy bounds? Let us start answering this question by
considering a universe undergoing decelerated expansion, that is  $H>0$, $\dot
H<0$. For entropy sources with $\alpha>-2$, going backwards in time, $H$ is
prevented by the restriction  $S^H \le  S_g^H$ from becoming too large. This
requires that at a certain moment in time $\dot H$ has reversed sign, or at least
vanished. GSL allows such a transition. Evolving from the past towards the future,
and looking at eq.~(\ref{ec}) we see that  a transition from an epoch of
accelerated expansion $H>0$, $\dot H >0$, to an epoch of decelerated expansion
$H>0$, $\dot H <0$, can occur without violation of GSL. But later we discuss a new
bound appearing in this situation when quantum effects are included.

For a contracting Universe with $H<0$, and if sources with $\alpha>-2$ exist, the
situation is more interesting. Let us check whether in an epoch of accelerated
contraction   $H<0$, $\dot H <0$, GSL is compatible with entropy bounds. If an
epoch of accelerated contraction lasts, it will inevitably run into a future
singularity, in conflict with  bound  $S^H \le  S_g^H$. This conflict could perhaps
have been prevented if at some moment in time the evolution had turned into
decelerated contraction with $H<0$, $\dot H > 0$. But a brief look at
eq.~(\ref{ec}), $\dot H \le -\frac{3}{3+\alpha} H^2$, shows that decelerated
contraction is not allowed by GSL.  The conclusion is that for the case of
accelerated contraction GSL and the entropy bound are not compatible.

To resolve the conflict between GSL and the entropy bound, we propose adding a
missing quantum entropy term $
 dS_{Quantum}=  - \mu dn_H,
$ where $\mu(a,H,\dot H,...)$ is a ``chemical potential" motivated by the following
heuristic argument. Specific quantum entropy is given by (\ref{quantums}), and we
consider for the moment one type of quantum fluctuations that preserves its
identity throughout the evolution. Changes in $S_q$ result from the well known
phenomenon of freezing and defreezing of quantum fluctuations. For example, quantum
modes whose wavelength is stretched by an accelerated cosmic expansion to the point
that it is larger than the horizon, become frozen (``exit the horizon"), and are
lost as dynamical modes, and conversely quantum modes whose wavelength shrinks
during a period of decelerated expansion (``reenter the horizon"), thaw and become
dynamical again. Taking into account this ``quantum leakage" of entropy, requires
that the first law should be modified as in open systems $TdS=dE+PdV-\mu dN$.

Consider a universe going through a period of decelerated expansion, containing
some quantum fluctuations which have reentered the horizon (for concreteness, it is
possible to  think about an isotropic background of gravitational waves). In this
case, physical momenta simply redshift, but since no new modes have reentered, and
since occupation numbers do not change by simple redshift, then within a fixed
comoving volume, entropy does not change. However, if there are some frozen
fluctuations outside the horizon ``waiting to reenter" then there will be a change
in quantum entropy, because the minimal comoving wave number of  dynamical modes
$k_{min}$, will decrease due to the expansion, $k_{min}(t+\delta t)<k_{min}(t)$.
The resulting change in quantum entropy, for a single physical degree of freedom,
is $
 \Delta s_{q}=\!\!\!
  \int\limits_{k_{min}(t+\delta t)}^{k_{min}(t)}\!\!\!
  k^2 dk \ln n_k,
  $
 and since $k_{min}(t)=a(t) H(t)$,
$
 \Delta S_{q}= \!\!\! \int\limits_{a(t+\delta t)H(t+\delta t)}^{a(t) H(t)}
  k^2 dk \ln n_k =
 -  \Delta (aH)^3 \ln n_{k=aH},
$ provided $\ln n_k$ is a smooth enough function. Therefore, for $\N$ physical DOF,
and since $n_H=(aH)^3$,
\begin{equation}
 d S_{q} = - \mu \N d n_H,
 \label{conjecture}
\end{equation}
where parameter $\mu$ is taken to be positive. Obviously, the result depends on the
spectrum $n_k$, but typical spectra are of the form $n_k\sim k^\beta$, and
therefore we may take as a reasonable approximation  $\ln n_k\sim constant$ for all
$\N$ physical DOF.

We adopt proposal (\ref{conjecture}) in  general,
\begin{eqnarray}
 dS &=& dS_{Classical}+dS_{Quantum} \nonumber \\
  &=&dn_H S^H+ n_H dS^H- \mu \N dn_H,
 \label{q2ndlaw}
\end{eqnarray}
where $S^H$ is the classical entropy within a cosmological horizon.
 In particular, for the case that $S^H$ is dominated by a single
 source $S^H=|H|^\alpha$,
\begin{equation}
 \left(3H+  3  \frac{{\dot H}}{H}\right) n_H (S^H-  \mu \N) +
 \alpha \frac{{\dot H}}{H} n_H S^H  \ge 0.
\label{qec}
\end{equation}

Quantum modified GSL (\ref{qec}) allows a transition from accelerated to
decelerated contraction. As a check, look at $H<0$, $\dot H=0$, in this case
modified GSL requires $ 3H (S^H-\mu \N) \ge 0, $ which, if $\mu \N \ge S^H $, is
allowed.  If the dominant form of entropy is indeed geometric entropy, the
transition from accelerated to decelerated contraction is allowed already at
$|H|\sim M_P/ \sqrt{\N}$. In models where $\N$ is a large number, such as grand
unified theories and string theory where it is expected to be  of the order of
1000, the transition  can occur  at a scale much below the Planck scale, at which
classical general relativity is conventionally expected to adequately describe
background evolution.

If we reconsider the transition from accelerated  to decelerated expansion and
require that (\ref{qec}) holds, we discover a new bound derived directly from GSL.
It is compatible with,  but not  relying on, the bound  $S^H \le  S_g^H$. Consider
the case in which $\dot H$ and $H$ are positive, or $H$ positive and $\dot H$
negative but $|\dot H|\ll H^2$, relevant to whether the transition is allowed by
GSL. In this case, (\ref{qec}) reduces to $S^H-\mu \N\ge 0$, that is, GSL puts a
lower bound on the classical entropy within the horizon. If geometric entropy is
the dominant source of entropy as expected,  GSL puts a lower bound on geometric
entropy $S_g^H\ge \mu \N$, which yields an upper bound on $H$,
\begin{equation}
 H\le \frac{M_P}{\sqrt{\N}}.
\label{upperH}
\end{equation}
The scale that appeared previously in the resolution of the conflict between
entropy bounds and GSL for a contracting universe has reappeared in (\ref{upperH}),
and remarkably, (\ref{upperH}) is the same bound obtained in \cite{Bek3} using
different arguments. Bound (\ref{upperH}) forbids a large class of singular
homogeneous, isotropic, spatially flat cosmologies by bounding the scale of
curvature for a such a universe.

\subsection{The Generalized Second Law in Pre-Big-Bang string cosmology}

String theory is  a consistent theory of quantum gravity, with the power to
describe high curvature regions of space-time \cite{Polchinski}, and as such we
could expect it to teach us about the fate of cosmological singularities, with the
expectation that singularities are smoothed and turned into brief epochs of high
curvature. However, many attempts to seduce an answer out of string theory
regarding cosmological singularities have failed so far in producing a conclusive
answer (see for example \cite{stringsing}). The reason is probably that most
technical advancements in string theory rely heavily on supersymmetry, but generic
time dependent solutions break all supersymmetries and therefore known methods are
less powerful when applied to cosmology.

We have focused \cite{GSLstring} on the two sources of entropy defined previously.
The first source is the geometric entropy $S_g$,  and the second source is quantum
entropy $S_{q}$. The entropy within a given horizon is $S^H$ and the total entropy
is given by $ S=n_H S^H$. We will ignore numerical factors, use units in which
$c=1$, $\hbar=1$, $G_{N}=e^{\phi}/16\pi$, $\phi$ being the dilaton, and discuss
only flat, homogeneous, and isotropic 4D string cosmologies in the so-called string
frame, in which the lowest order effective action is
 ${\cal S}_{\rm LO}=\int d^4 x \sqrt{-g} e^{-\phi}
 \left[R + \left(\partial\phi\right)^2\right]$. Obviously the discussion can be
 generalized in a straight forward manner to higher $D$.

In ordinary cosmology, geometric entropy within a Hubble volume is given by its
area $S_g^H=H^{-2} G_N^{-1}$, and therefore specific geometric entropy is given by
 $s_{g}=|H| G_{N}^{-1}$ \cite{GSL}. A possible expression for
specific geometric entropy in string cosmology is obtained by substituting
$G_{N}=e^{\phi}$, leading to
\begin{equation}\label{sst}
s_{g}=|H| e^{-\phi}.
\end{equation}
Reassurance that $s_g$ is indeed given by (\ref{sst}) is provided by the following
observation. The action ${\cal S}_{\rm LO}$ can be expressed in a $(3+1)$ covariant
form, using the 3-metric $g_{ij}$, the extrinsic curvature $K_{ij}$, considering
only vanishing $3-$Ricci scalar and homogeneous dilaton, $ {\cal S}_{\rm LO}=\int
d^3 x dt \sqrt{g_{ij}} e^{-\phi}
 \left[ - 3 K_{ij}K^{ij} - 2 g^{ij}\partial_t{K_{ij}}
 + K^2 - (\partial_t {\phi})^2 \right].
$ Now, ${\cal S}_{\rm LO}$  is invariant under the symmetry transformation $
g_{ij}\rightarrow e^{2 \lambda}g_{ij}$,
 $\phi \rightarrow \phi +3\lambda$, for an arbitrary time
 dependent $\lambda$.
From the variation of the action $\delta {\cal S}=\int d^3 x dt \sqrt{g_{ij}}
e^{-\phi} 4 K \dot{\lambda}$, we may read off the current and conserved charge $Q=4
a^3 e^{-\phi} K$. The  symmetry is exact in the flat homogeneous case, and it seems
plausible that it is a good symmetry even when  $\alpha '$ corrections are present
\cite{Gasperini:1997fu}. With definition (\ref{sst}), the total geometric entropy $
S_{g}=a^3 |H| e^{-\phi}, $ is proportional to the corresponding conserved charge.
Adiabatic evolution, determined by $\partial_t S_g=0$, leads to a familiar
equation, $ \frac{\dot{H}}{H}-\dot{\phi}+3H= 0, $ satisfied by the $(\pm)$ vacuum
branches of PBB string cosmology.

Quantum entropy for a single field in string cosmology is, as in \cite{BMP,gg,GSL},
given by
\begin{equation} \label{squantum}
 s_{q}=\int_{k_{min}}^{k_{max}} d^3k f(k)\, ,
\end{equation}
where for large occupation numbers $f(k)\simeq \ln n_k$. The ultraviolet cutoff
$k_{max}$ is assumed to remain constant at the string scale. The infrared cutoff
$k_{min}$ is determined by the perturbation equation $
 \psi_{k_{c}}''+\left(k_{c}^2- \frac{\sqrt{s(\eta)} ''}{\sqrt{s(\eta)}}\right)
\psi_{k_{c}}=0, $ where $\eta$ is conformal time $'=\partial_\eta$, and $k_{c}$ is
the comoving momentum related to physical momentum $k(\eta)$ as $k_{c}=a(\eta)
k(\eta)$. Modes for which $k_{c}^2 \leq \frac{\sqrt{s}''}{\sqrt{s}}$ are ``frozen",
and are lost as dynamical modes. The ``pump field" $s(\eta)=a^{2m} e^{\ell\phi}$,
depends on the background evolution and on the spin and dilaton coupling of various
fields. We are interested in solutions for which $a'/a\sim \phi' \sim 1/\eta$, and
therefore, for all particles $\frac{\sqrt{s}''}{\sqrt{s}}\sim 1/\eta^2$. It follows
that $k_{min}\sim H$. In other phases of cosmological evolution our assumption does
not necessarily hold, but in standard radiation domination (RD) with frozen dilaton
all modes reenter the horizon. Using the reasonable approximation $f(k)\sim$
constant, we obtain, as in \cite{GSL},
\begin{equation} \label{dtsquan}
 \Delta S_{q}\simeq-\mu \Delta n_H.
\end{equation}
Parameter $\mu$ is positive, and  in many cases  proportional to the number of
species of particles, taking into account all DOF of the system, perturbative and
non-perturbative. The main contribution to $\mu$ comes from light DOF and therefore
if some non-perturbative objects, such as D branes become light they will make a
substantial contribution to $\mu$.

We now turn to the  generalized second law of thermodynamics, taking into account
geometric and quantum entropy. Enforcing $dS\ge 0$, and in particular,
 $\partial_t S= \partial_t S_g + \partial_t S_q\ge 0$,
 leads to an important inequality,
\begin{equation} \label{eq:gsl}
 \left(H^{-2} e^{-\phi}-\mu\right)\partial_t n_H+
 n_H \partial_t \left(H^{-2} e^{-\phi}\right) \ge 0.
\end{equation}
When quantum entropy is negligible compared to geometric entropy, GSL
(\ref{eq:gsl}) leads to
\begin{equation}\label{dotphibound}
\dot{\phi}\leq \frac{\dot{H}}{H}+3H,
\end{equation}
yielding a bound on $\dot{\phi}$, and therefore on dilaton kinetic energy,  for a
given $H$, $\dot{H}$. Bound (\ref{dotphibound}) was first obtained in \cite{HEB},
and interpreted as following from a saturated HEB.

When quantum entropy becomes relevant we obtain another bound. We are interested in
a situation in which the universe expands, $H>0$, and $\phi$ and $H$ are
non-decreasing, and therefore
 $\partial_t \left(H^{-2} e^{-\phi}\right)\le 0$ and
 $\partial_t n_H>0$. A necessary condition for GSL to hold is that
\begin{equation} \label{eq:hbound}
 H^{2} \le \frac{ e^{-\phi}}{\mu},
\end{equation}
bounding total geometric entropy $H e^{-\phi}\leq \frac{e^{-\frac{3}{2}
\phi}}{\sqrt{\mu}}$. A bound similar to (\ref{eq:hbound}) was obtained in
\cite{HEB} by considering entropy of reentering quantum fluctuations. We stress
that to be useful in analysis of cosmological singularities (\ref{eq:hbound}) has
to be considered for perturbations that exit the horizon. If the condition
(\ref{eq:hbound}) is satisfied then the cosmological evolution always allows a
self-consistent description using the low energy effective action approach.

It is not apriori clear that the form of GSL and entropy sources remains unchanged
when curvature becomes large, in fact, we may expect higher order corrections to
appear. For example, the conserved charge of the scaling symmetry of the action
will depend in general on higher order curvature corrections. Nevertheless, in the
following we will assume that  specific geometric entropy is given by
eq.~(\ref{sst}), without higher order corrections, and try to verify that, for some
reason yet to be understood, there are no higher order corrections to
eq.~(\ref{sst}). Our results are consistent with this assumption.

We turn now to apply our general analysis to the PBB string cosmology scenario, in
which the universe starts from a state of very small curvature and string coupling
and then undergoes a long phase of dilaton-driven inflation (DDI), joining smoothly
at later times  standard RD cosmology,  giving rise to a singularity free
inflationary cosmology. The high curvature phase joining DDI and RD phases is
identified with the `big bang' of standard cosmology. A key issue confronting this
scenario is whether, and under what conditions, can the graceful exit transition
from DDI to RD be completed \cite{Brustein:1994kw}. In particular, it was argued
that curvature is bounded by an algebraic fixed point behaviour  when both $H$ and
$\dot\phi$ are constants and the universe is in a linear-dilaton deSitter space
\cite{Gasperini:1997fu}, and coupling is bounded by quantum corrections
\cite{Brustein:1997ny,Foffa:1999dv}. But it became clear that another general
theoretical ingredient is missing, and we propose that GSL is that missing
ingredient.

We have studied numerically  examples of PBB string cosmologies to verify that the
overall picture we suggest is valid in cases that can be analyzed explicitly. We
first consider, as in \cite{Gasperini:1997fu,Brustein:1999yq}, $\alpha'$
corrections to the lowest order string effective action,
\begin{equation}\label{eq2}
{\cal S}=\frac{1}{16 \pi \alpha '}\int d^4 x \sqrt{-g} e^{-\phi}\left[R +
\left(\partial\phi\right)^2 + \frac{1}{2}{\cal L}_{\alpha '}\right]\, ,
\end{equation}
where
\begin{eqnarray}\label{alfa}
 {\cal L}_{\alpha '}=k \alpha' \Biggl[&& \frac{1}{2}R^2_{GB} + A
\left(\partial\phi\right)^4 +
D \partial^2\phi\left(\partial\phi\right)^2 \nonumber \\
&& +C\left(R^{\mu\nu}-\frac{1}{2}g^{\mu\nu}R\right)
\partial_{\mu}\phi\partial_{\nu}\phi\Biggr],
\end{eqnarray}
with $C=-(2 A + 2D +1)$, is the most general form of four derivative corrections
that lead to equations of motion with at most  second (time) derivatives. The
rationale for this choice was explained in \cite{Brustein:1999yq}.  $k$ is a
numerical factor depending on the type of string theory. Action (\ref{eq2}) leads
to equations of motion, $ - 3 H^2 + \dot{\bar{\phi}}^2- \bar{\rho}=0 $, $
\bar{\sigma} - 2 \dot{H} + 2 H \dot{\bar{\phi}}=0 $, $ \bar{\lambda} - 3 H^2 -
\dot{\bar{\phi}}^2 + 2 \ddot{\bar{\phi}}=0 $, where $\bar{\rho}$, $\bar{\lambda}$,
$\bar{\sigma}$ are effective sources parameterizing the contribution of $\alpha '$
corrections \cite{Brustein:1999yq}. Parameters $A$ and $D$ should have been
determined  by string theory, however, at the moment, it is not possible to
calculate them in general. If $A$, $D$  were determined we could just use the
results and check whether their generic cosmological solutions are non-singular,
but since $A$, $D$ are unavailable at the moment, we turn to GSL to restrict them.

\begin{figure}
\centering
\includegraphics[height=4cm]{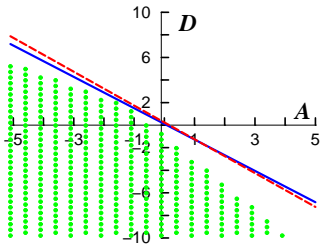}
\caption{ {\label{fig1r}} Two lines, separating actions whose generic solutions
``turn the right way'' at the early stages of evolution (red-dashed),
 and actions whose generic solutions satisfy classical GSL
while close to the $(+)$ branch vacuum (blue-solid). The dots represent  $(A,D)$
values whose generic solutions reach a fixed point, and are all in the "allowed"
region.}
\end{figure}

First, we look at the initial stages of the evolution when the string coupling and
$H$ are very small. We find that not all the values of the parameters $A$, $D$ are
allowed by GSL. The condition $\bar{\sigma}\geq0$, which  is equivalent to GSL on
generic solutions at the very early stage of the evolution, if the only relevant
form of entropy is geometric entropy, leads to the following condition  on  $A$,
$D$ (first obtained by R. Madden \cite{dick}), $ 40.05 A + 28.86 D \leq 7.253. $
The values of $A$, $D$ which satisfy this inequality are labeled ``allowed", and
the rest are ``forbidden". In \cite{Brustein:1999yq} a condition that $\alpha'$
corrections are such that solutions start to turn towards a fixed point at the very
early stages of their evolution was found $ 61.1768 A + 40.8475 D \leq 16.083$, and
such solutions were labeled ``turning the right way". Both conditions  are
displayed in Fig.~\ref{fig1r}. They select almost the same region of $(A,D)$ space,
a gratifying result, GSL ``forbids" actions whose generic solutions are singular
and do not reach a fixed point.

\begin{figure}
\centering
\includegraphics[height=4cm]{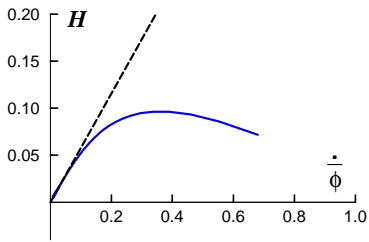}
\caption{ {\label{fig2r}} Typical solution that
 ``turns the wrong way". The dashed line is the $(+)$ branch vacuum.}
\end{figure}

We further observe that generic solutions which ``turn the  wrong way" at the early
stages of their evolution continue their course in a way similar to the solution
presented in Fig.~\ref{fig2r}. We find numerically that at a certain moment in time
$H$ starts to decrease, at that point $\dot{H}=0$ and particle production effects
are still extremely weak, and therefore (\ref{dotphibound}) is the relevant bound,
but (\ref{dotphibound}) is certainly violated.

We have scanned the $(A,D)$ plane  to check whether a generic solution that reaches
a fixed point respects GSL throughout the whole evolution, and conversely, whether
a generic solution obeying GSL evolves towards a fixed point. The results are shown
in Fig.~\ref{fig1r}, clearly, the ``forbidden" region does not contain actions
whose generic solutions go to fixed points. Nevertheless, there are some $(A,D)$
values located in the small wedges near the bounding lines, for which the
corresponding solutions always satisfy (\ref{dotphibound}), but do not reach a
fixed point, and are singular. This happens because they meet a cusp singularity.
Consistency requires adding higher order $\alpha'$ corrections when cusp
singularities are approached, which we will not attempt here.

If particle production effects are strong, the quantum part of GSL adds  bound
(\ref{eq:hbound}), which  adds another ``forbidden" region in the
$(H,\dot{\bar{\phi}})$ plane, the region above a straight line parallel to the
$\dot{\bar{\phi}}$ axis. The quantum part of GSL has therefore a significant impact
on corrections to the effective action.   On a fixed point $\phi$ is still
increasing, and therefore the bounding line described by (\ref{eq:hbound}) is
moving downwards, and when the critical line moves below the fixed point,  GSL is
violated. This means that when a certain critical value of the coupling $e^{\phi}$
is reached, the solution can no longer stay on the fixed point, and it must move
away towards an exit. One way this can happen is if quantum corrections, perhaps of
the type discussed in \cite{Brustein:1997ny,Foffa:1999dv} exist.

\begin{figure}
\centering
\includegraphics[height=4cm]{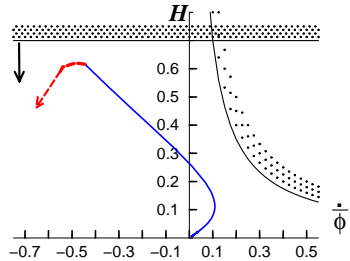}
\caption{ {\label{fig4r}} Graceful exit enforced by GSL on generic solutions. The
horizontal line is bound (\ref{eq:hbound}) and the curve on the right is bound
(\ref{dotphibound}),  shaded regions indicate  GSL violation.}
\end{figure}

The full GSL therefore forces actions to have generic solutions that are
non-singular, classical GSL bounds
 dilaton kinetic energy and quantum  GSL bounds $H$
and therefore, at a certain moment of the evolution  $\dot{H}$ must vanish (at
least asymptotically), and then curvature is bounded. If  cusp singularities are
removed by adding higher order corrections, as might be expected,  we can apply GSL
with similar conclusions also in this case. A schematic graceful exit enforced by
GSL is shown in Fig.~\ref{fig4r}.  Our result indicate that if we impose GSL in
addition to equations of motion then  non-singular PBB string cosmology is quite
generic.

\section{Area entropy, entanglement entropy and entropy bounds}
\label{sec:5}

Classical General Relativity predicts space-times with event horizons and other
causal boundaries, such as apparent horizons, cosmological horizons and
acceleration horizons. Observers in space-times with causal boundaries can see very
different physics, as demonstrated by comparing the static observer at infinity and
a freely falling observer in the Schwarzchild geometry. For the first, the horizon
is a very special place: energies of particles diverge and space-time seems to end
there, while for a freely falling observer the horizon and its vicinity do not look
special at all. In cosmological space-times with causal boundaries the situation is
similar. The existence of causal boundaries is determined by the large scale
properties of space-time, and hence is intrinsically a non-local concept. In
cosmology, for example, it is hard for a local observer to determine whether the
space-time is de Sitter space that has a cosmological event horizon, or a
Robertson-Walker space which looks approximately de Sitter.

The interpretation of the thermodynamic properties of BH's and whether they
originate from some underlying, more fundamental, statistical mechanics remains
unclear, in spite of the intense efforts and the progress that has been achieved
over the last 30 years since the discovery by Bekenstein \cite{gslbh}. Quantum
field theory (QFT) in the fixed background of space-times with horizons is a key
element in the quantitative understanding of the statistical mechanics of BH's. QFT
in such background  has several interesting and well-known features. The quantum
vacuum states associated with different observers can be very different from each
other, leading to strong particle production effects:  the Hawking effect and the
Unruh effect. In addition, the appearance of large blue shifts of quantum modes
near the horizon lead to the trans-planckian  problem \cite{Jacobson:2003vx}. The
proposed resolutions include the brick-wall model
\cite{'tHooft:1984re,'tHooft:1996tq} and the stretched-horizon
\cite{Susskind:1993if} idea. The entropy and thermodynamics are also observer
dependent, as demonstrated by the classic comparison between the Rindler and
Minkowski space observers in the Minkowski vacuum. The accelerated observer sees a
truly thermal state, while for the Minkowski observer the temperature vanishes. The
tension between the possibility of evaluating the entropy and other thermodynamic
quantities in the semiclassical approximation and their observer dependence and
hence their sensitivity to physics at the highest energy scales is intriguing and
is not yet resolved.

My current point of view about the physics of space-times with causal boundaries is
the entanglement point of view. I believe that the statistical properties of such
space-times arise because classical observers in them have access only to a part of
the whole quantum state. When a system is in a pure state, but one cannot access
the complete quantum system, and a measurement is performed, one is instructed by
the rules of quantum mechanics to trace over the classically inaccessible DOF. This
leads to a natural framework for interpreting the physics of spaces with causal
boundaries: that it is described by the density matrix which results from tracing
over the inaccessible DOF.  In the context of BH's the idea was first proposed by
't Hooft \cite{'tHooft:1996tq}, and by Sorkin and collaborators
\cite{Bombelli:1986rw}, and then extended and elaborated by Srednicki
\cite{Srednicki:1993im} and others.

The entanglement approach considers the fundamental physical objects describing the
physics of space-times with causal boundaries to be their global quantum state and
the unitary evolution operator. The entanglement approach has several obvious
advantages: it leads naturally to area-law entropy, it can incorporate the observer
dependence of BH thermodynamics and of the thermodynamics of cosmological
space-times with causal boundaries. It can naturally accommodate the geometric and
quantum entropies. The first resulting from the entanglement entropy of short
wavelength fluctuations and the second resulting from the entanglement entropy  of
fluctuations whose wavelength is larger than the causal connection scale. This
interpretation is also automatically compatible with entropy bounds and the GSL as
long as the evolution equations are ``physical" because from a global point of view
it is clear that nothing special occurs when a horizon develops. Obviously, there
are also some unresolved issues that need to be understood better in this context.

The space-times that are traditionally used to explore the entanglement point of
view are spaces with bifurcating Killing horizons such as the eternal Schwarzschild
BH or Rindler space. Israel \cite{Israel:1976ur} has shown that the quantum Hilbert
space of fields in space-times with bifurcating Killing horizons has a product
structure that is isomorphic to the product structure that arises in thermofield
dynamics \cite{Takahasi:1974zn}. In thermofield dynamics one formally doubles the
Hilbert space and evaluates quantum expectation values in the thermofield double
pure state in order to evaluate expectation values in a thermal state of the
original system. In this context the entropy is the entanglement entropy that is
obtained from tracing over one of the two spaces.

One of the main unresolved issues confronting the entanglement interpretation is
the ultraviolet (UV) divergence of entanglement entropy and other entanglement
correlation functions near the horizon, and its dependence on the number of fields
\cite{Wald:1999vt,Marolf:2003bb,Marolf:2004et}. Another issue concerns space-times
that do not have non-degenerate bifurcating Killing horizons. For such spaces, it
is unclear what is entangled with what, since some of the regions of the extended
space-time are missing.

The entanglement point of view has been discussed in the AdS-CFT context by
Maldacena \cite{Maldacena:2001kr} who studied eternal BH's in AdS.  In 4D, the
space has two boundaries that are topologically $S^2 \times S^1$, the dual FT
consists of two CFTs ``living on the boundary". The product theory in the TFD state
defines the string theory in the bulk, whose low energy limit is the AdS-BH. The FT
side is completely well-defined, and its thermodynamics can obviously be
interpreted as entanglement thermodynamics. The low energy state in the bulk is the
Hartle-Hawking vacuum. The entanglement point of view suggests the following
perspective. Suppose that the universe is in a pure state and that it evolves
unitarily. Then the entropy of any sub-system of it is entirely in the eyes of the
beholder: a particular classical observer.

We have shown \cite{Brustein:2005vx} that the entropy resulting from the counting
of microstates of non-extremal BH's using field theory duals of string theories can
be interpreted as arising from entanglement. The conditions for making such an
interpretation consistent were determined. First, we have interpreted the entropy
and thermodynamics of spacetimes with non degenerate, bifurcating Killing horizons
as arising from entanglement.  We have used a path integral method to define the
Hartle-Hawking vacuum state in such spacetimes, and reveal explicitly its entangled
nature and its relation to the geometry.   If string theory on such spacetimes has
a field theory dual, then, in the low-energy, weak coupling limit, the field theory
state that is dual to the Hartle-Hawking state is a thermofield double state. This
allowed us to compare the entanglement entropy to the entropy of the field theory
dual, and thus to the Bekenstein-Hawking entropy of the BH.

To further understand the nature of the time evolution of subsystems in this
context we have considered \cite{Brustein:2006wp} a collapsing relativistic
spherical shell in a free quantum field theory. Once the center of the wavefunction
of the shell passes a certain radius $r_s$, the degrees of freedom inside $r_s$ are
traced over. We have found that an observer outside this region will determine that
the evolution of the system is non-unitary. The non-unitary evolution occurs only
when the wavefunction is in the process of crossing the boundary and the amount of
non-unitarity is proportional to the area of the boundary.

\section{Acknowledgments}
\label{sec:6}

I would like to thank all the collaborators who participated in the research that
is summarized and reviewed in this article. First, I would like to thank Gabriele
Veneziano for interesting me in this subject and for collaboration in several
related projects. I would like to thank David Eichler, Marty Einhorn,  Stefano
Foffa, Dick Madden, David Oaknin,  Avi Mayo, Riccardo Sturani and Amos Yarom for
fruitful collaborations whose results are presented in this article.

%
%
%
%
%
%
%
%

\end{document}